\documentclass{aa}

\usepackage{rotating}
\usepackage{graphics}
\usepackage{psfig}
\usepackage{graphicx}
\usepackage{txfonts}

\def \hi {H\,{\sc i~}}

\def\NH{$N_{\rm HI}$}
\def\kms{km\,s$^{-1}$}
\def\deg{\hbox{$^\circ$}}

\begin{document}

\title{The WSRT wide-field \hi survey}
\subtitle{I. The background galaxy sample}
\titlerunning{Background galaxies in the WSRT wide-field survey}
\authorrunning{Braun, Thilker \& Walterbos}

\author{
  Robert Braun\inst{1} \and
  David Thilker\inst{2} \and
  Rene A.\,M. Walterbos\inst{3}
}

\institute{
  ASTRON,
  P.O. Box 2,
  7990 AA Dwingeloo,
  The Netherlands \and
Department of Physics and Astronomy,
Johns Hopkins University, 
3400 N. Charles St., 
Baltimore MD 21218-2695, U.S.A. \and
Department of Astronomy,
New Mexico State University,
Box 30001, MSC 4500,
Las Cruces NM 88003, U.S.A. }

\date{Received mmddyy / Accepted mmddyy}

\offprints{R. Braun,
\email{rbraun@astron.nl}}

\abstract{ We have used the Westerbork array to carry out an unbiased
wide-field survey for \hi emission features, achieving an {\sc rms}
sensitivity of about 18~mJy/Beam at a velocity resolution of 17~\kms\ 
over 1800~deg$^2$ and between $-1000~<~$V$_{Hel}~<~+6500$~\kms. The
primary data consists of auto-correlation spectra with an effective
angular resolution of 49$^\prime$ FWHM, although cross-correlation data
were also acquired. The survey region is centered approximately on the
position of Messier~31 and is Nyquist-sampled over 60$\times$30$^\circ$
in R.A.$\times$Dec. More than 100 distinct features are detected at
high significance in each of the two velocity regimes (negative and
positive LGSR velocities). In this paper we present the results for our
\hi detections of external galaxies at positive LGSR velocity. We
detect 155 external galaxies in excess of 8$\sigma$ in integrated \hi
flux density. Plausible optical associations are found within a
30$^\prime$ search radius for all but one of our \hi detections in DSS
images, although several are not previously cataloged or do not have
published red-shift determinations. Our detection without a DSS
association is at low galactic latitude. Twenty-three of our objects are
detected in \hi for the first time. We classify almost half of our
detections as ``confused'', since one or more companions is cataloged
within a radius of 30$^\prime$ and a velocity interval of 400~\kms.  We
identify a handful of instances of significant positional offsets
exceeding 10~kpc of unconfused optical galaxies with the associated \hi
centroid, possibly indicative of severe tidal distortions or
uncataloged gas-rich companions. A possible trend is found for an
excess of detected \hi flux in unconfused galaxies within our large
survey beam relative to that detected previously in smaller telescope
beams, both as function of increasing distance and increasing gas
mass. This may be an indication for a diffuse gaseous component on
100~kpc scales in the environment of massive galaxies or a population
of uncataloged low mass companions. We use our
galaxy sample to estimate the \hi mass function from our survey
volume. Good agreement is found with the HIPASS BGC results, but only
after explicit correction for galaxy density variations with distance.

 }

\maketitle

\section{Introduction}

Unbiased wide-field surveys are an indispensible means for determining
the physical content of our extended environment. The SDSS (Sloan
Digital Sky Survey (York et al. \cite{york00})) is a prime example of
the way in which such work is providing new insights at optical
wavelengths into both the nearby and distant universe. At radio
frequencies there have been a number of wide-field surveys for both
continuum sources and, to a lesser extent, emitters and absorbers in
specific spectral lines. The HIPASS survey (Barnes et
al. \cite{barn01}) marks an important milestone in achieving high
sensitivity to \hi emission over more than half of the sky (given it's
ongoing extension of the Declination coverage from 0 to
+25\deg). HIPASS has provided a deep inventory of both negative and
positive LGSR (Local Group Standard of Rest) velocity \hi emission
features. The negative velocity features (Putman et al. \cite{putm02})
are primarily associated in some way with the Galaxy and other Local
Group objects, while the positive velocity features are primarily
associated with moderately nearby ($<$ 100~Mpc) external galaxies
(eg. Kilborn et al. \cite{kilb02}). A northern hemisphere counterpart
to the HIPASS survey is now underway in the form of HIJASS (\hi Jodrell
All Sky Survey, Lang et al. \cite{lang03}).

An interesting component of the negative velocity \hi sky are the
so-called compact high velocity clouds, CHVCs (Braun \& Burton
\cite{brau99}), which are isolated in position and velocity from the
more extended high velocity \hi complexes down to column densities
below about \NH~=~1.5$\times10^{18}$cm$^{-2}$. The suggestion has been
made that these objects may be the most distant component of the high
velocity cloud phenomenon, perhaps extending to 100's of kpc from their
host galaxies. A critical prediction of this scenario (De Heij et
al. \cite{dehe02}) is that a large population of faint CHVCs should be
detected in the vicinity of M31 (at declination +40\deg) if enough
sensitivity were available. While current observational data are 
consistent with this scenario, they are severely limited by the modest
point source sensitivity available at northern declinations (within the
Leiden/Dwingeloo Survey (Hartmann \& Burton \cite{hart97})) which is
almost an order of magnitude poorer than that of HIPASS in the south.

We have undertaken a moderately sensitive large-area \hi survey both to
test for the predicted population of faint CHVCs near M31 as well as to
carry out an unbiased search for \hi emission associated with
background galaxies.  We have achieved an {\sc rms} sensitivity of
about 18~mJy/Beam at a velocity resolution of 17~\kms\ over
1800~deg$^2$ and between $-1000~<~$V$_{Hel}~<~+6500$~\kms. The
corresponding {\sc rms} column density sensitivity for emission filling
the 3000$\times$2800 arcsec effective beam area is about
4$\times10^{16}$cm$^{-2}$ over 17~\kms. For comparison, the HIPASS
survey has achieved an {\sc rms} of about 14~mJy/Beam at a velocity
resolution of 18~\kms, yielding a slightly superior flux
sensitivity. On the other hand, the column density sensitivity for
emission filling our larger beam exceeds that of HIPASS by almost an
order of magnitude. Since the linear {\sc FWHM} diameter of our
survey beam varies from about 10~kpc at a distance of 
0.7~Mpc to more than 1~Mpc at 75~Mpc, it is only at Local Group
distances that the condition of beam filling is likely to be achieved.
Compared to the Leiden/Dwingeloo Survey, we achieve
an order of magnitude improvement in both flux density and brightness
sensitivity.  We detect more than 100 distinct features at high
significance in each of the two velocity regimes (negative and positive
LGSR velocities).  In this paper we will describe the survey
observations and data reduction procedures in
\S\,\ref{sec:observations}, followed by a presentation of the results
for our \hi detections of external galaxies in \S\,\ref{sec:results}
and closing with a brief discussion of these results in
\S\,\ref{sec:discussion}. Our results at negative LGSR velocities will
be presented in a companion paper.

\section{Observations and Data Reduction}
\label{sec:observations}

\subsection{Survey Strategy}

Our survey area was defined to have an extent of 60$\times$30 true
degrees oriented in $\alpha_{2000}\times\delta_{2000}$ and centered on
($\alpha_{2000},\delta_{2000}$)~=~(10\deg,35\deg), about 5\deg south of
the M31 nuclear position. Data were acquired in a drift-scan mode,
whereby the 25~m telescopes of the WSRT array were kept stationary at a
specified start position and the sky drifted past at the earth-rotation
rate.  Each telescope beamwidth is about 35 arcmin FWHM at an observing
frequency of 1410 MHz. The fourteeen telescopes of the array were split
into two sub-arrays of seven telescopes each. The two sub-arrays were
pointed at declinations offset from one another by 15 arcmin, in order
to achieve Nyquist-sampled declination coverage of the survey area in half
the time that would otherwise be required. The recorded data were
averaged over 60 sec, corresponding to an angular drift of about 15
arcmin of right ascension, to yield Nyquist-sampling in the scan
direction of the telescope beam.

Although the primary objective of the survey was acquisition of
auto-correlation data, it was also desirable to acquire
cross-correlation data simultaneously for the two sub-arrays of seven
telescopes which observed the same set of positions on the sky. To this
end, electronic tracking was employed during each 60 second integration
directed at the sequence of central positions that was sweeping through
the telescope beam at the earth-rotation rate. The two sub-arrays were each
composed of six telescopes with short relative spacings (betweem 36 and
144 m) and a seventh telescope at a larger separation (of about 1.5
km). The duration of the drift-scan observations varied with
declination from about 4.3 to 6.2 hours. A typical observing sequence
consisted of a standard observation of a primary calibration source
(3C48 or 3C286), a dual sub-array drift-scan observation and in some
cases a second dual sub-array drift-scan observation. Each such session
provided the survey data for a strip of either 60$\times$0.5 or
60$\times$1 true degrees. Thirty of the ``double'' sessions, lasting
some 320 hours, could in principle provide the complete survey
coverage.

In practise, the observations were distributed over some 52 sessions in
the period 2002/09/04 to 2002/11/16. An effort was made to acquire the
drift-scan data only after local sunset and before local sunrise to
minimize solar interference. This was largely successful, with only a
few hours of data showing the effects of the sun above the horizon. An
effort was also made to insure that the drift-scan data was only
acquired at moderately high elevations, both to eliminate the
possibility of inter-telescope shadowing and to optimize the system
temperature. Essentially all observations were done at elevations
between 45 and 85 deg, for which the system temperature variations are
observed to be less than 1~K, corresponding to about 3\%. Repeat
coverage of a number of scans was obtained in cases where instrumental
failure or severe interference led to a significant increase in the
noise level.

Data was acquired in two 20~MHz IF bands centered at 1416 and 1398
MHz. The 18~MHz spacing of the two bands was chosen to provide a
contiguous velocity coverage at a uniform nominal sensitivity.  All
auto- and cross-correlations were recorded for both linear
polarizations in 512 uniformly weighted spectral channels across each
20~MHz band. A hanning smoothing was applied after the fact to minimize
the spectral side-lobes of interference, yielding a spectral resolution
of 78.125~kHz, corresponding to about 16.6 and 16.8 \kms\  in the two
bands.

\subsection{Data Reduction}

The drift-scan data for each sub-array was inspected and flagged in
Classic AIPS using the SPFLG utility. Any questionable features
appearing in the (time,frequency) display of each auto-correlation
baseline were critically compared amongst the 14 independent estimates
(7 telescopes and 2 polarizations) that were available. Any features
which were not reproduced in the other simultaneous spectra (from
telescopes seperated by as much as 2~km) were flagged. This allowed
quite effective discrimination against interference. 

Absolute flux calibration of both the auto- and cross-correlation data
was provided by the observed mean cross-correlation coefficient
measured for the standard calibration sources (3C48 or 3C286) of known
flux density. The measured ratio of flux density to correlation
coefficient averaged over all 14 telescopes and 2 polarizations was
300$\pm$10~Jy/Beam. Although there are variations (typically less than
about 10\%) amongst the 28 independent receiver systems, the average
gain of the system (at this frequency) has remained constant at the
quoted value over a period of several years to better than 5 percent.
The calibrator observation of each observing session was used both to
determine the average gain value appropriate for the auto-correlations
as well as providing phase and gain solutions appropriate for the
calibration of the cross-correlation drift-scan data in that session.

Two different methods were employed to generate data-cubes of the
auto-correlation data. The first method employed a local robust average
of a 30 minute sliding window to estimate the band-pass as a function
of time and a 850~\kms\ sliding window to estimate the continuum level
as a function of frequency. Only those values between the first and
third quartiles were included in these averages, making them moderately
robust to outliers, including \hi emission features, in the data. This
method could be applied blindly and produced the most uniform noise
characteristics in the resulting cube. As such, it was well-suited for
the automated detection of faint sources. However, moderately bright
sources of \hi emission that were extended either spatially or in
velocity produced localized negative artifacts. The best results under
these circumstances were obtained with the more complicated procedure
outlined below: 1) a quadratic baseline in time was fit to the entire
drift-scan (of 4 to 6 hour duration) and divided out after masking out
any localized regions of emission, 2) a constant offset was determined
and divided out of each frequency spectrum after masking out any
regions of line emission (including the extended emission from the
Galaxy), 3) a quadratic baseline in frequency was fit and subtracted
from the lower half (the first 10 MHz) of each frequency spectrum after
masking out any regions of line emission, 4) an alternative baseline
solution (primarily for the upper 10~MHz of each band) was derived from
a boxcar smoothed spectrum with 6 MHz box-width and subtracted from
each frequency spectrum after masking out any regions of line emission,
5) a cubic baseline in time was fit and subtracted out of each
drift-scan after masking out any localized regions of emission, 6) all
unflagged data for each position and frequency were averaged, 7) the
entire process (steps (1) through (6) above) was repeated using an
updated mask to isolate regions of significant emission from the
baseline determination.

The rationale for each step noted above was the folowing; step (1) was
intended to compensate for long timescale variations in the basic
bandpass shape, step (2) for the possible contributions of bright
continuum sources to the system temperature as function of time, steps
(3) and (4) for residual corrections to the bandpass shape on short
timescales (where it was found empirically that the two different methods
gave somewhat superior results in the two halves of the band) and step
(5) for residual corrections to the bandpass shape on intermediate
timescales. 

Although the survey strategy was designed to provide the most stable
possible bandpass response, this proved to be somewhat
disappointing. During the course of the observing campaign it was
established that there were systematic variations in the bandpass shape
at the level of about 1:1000 which were closely correlated with small
variations in ambient temperature in the vicinity of the IF system
electronics on timescales of 0.5 to 5 hours. Steps were taken to
stabilize the airflow and ambient temperature which led to a
substantial improvement in baseline stability. Even so, the remaining
fluctuations are still a limiting factor to the final auto-correlation
sensitivity, at least in the upper one third of each 20 MHz band, where
they were most severe.

The drift-scan data were resampled in frequency to convert from the
fixed geocentric frequencies of each observing date to heliocentric
radial velocities at each observed position. A Gaussian smoothing with
1800 true arcsec FWHM was applied in RA before combining the 120
drift-scans into a data-cube for each 20 MHz band. A spatial resampling
was then carried out to convert from the rectangular ($\alpha,\delta$)
of the acquired data to a projected ($\alpha,\delta$)
geometry. Finally, a spatial convolution with a 1800$\times$900 arcsec
FWHM Gaussian with PA~=~0\deg was applied to introduce the desired
degree of spatial correlation in the result.

The average primary beam of the WSRT array at an observing frequency of
1400~MHz has been determined with a holographic measurement (and can be
found depicted at
http://www.astron.nl/wsrt/wsrtGuide/WSRT21BEAM.PS). The central
positive lobe has an area of 1535~arcmin$^2$ and is moderately well-fit
by a Gaussian with 2235 arcsec FWHM. Peak sidelobe levels are about
0.1\%.  The same sequence of spatial smoothings described above, first
a 900'' box-car in RA to simulate the drift-scan, then a 1800''
Gaussian in RA and then a 900$\times$1800'' Gaussian in
($\alpha,\delta$) was applied to a digital representation of the
primary beam.  This allowed estimation of the effective beam size for
the survey, which can be approximated by a Gaussian of 3020$\times$2810
arcsec at PA~=~90\deg, as well as the beam dilution factor needed to
preserve the absolute flux scale in units of Jy/Beam.

\begin{figure}
 \resizebox{\hsize}{!}{\includegraphics{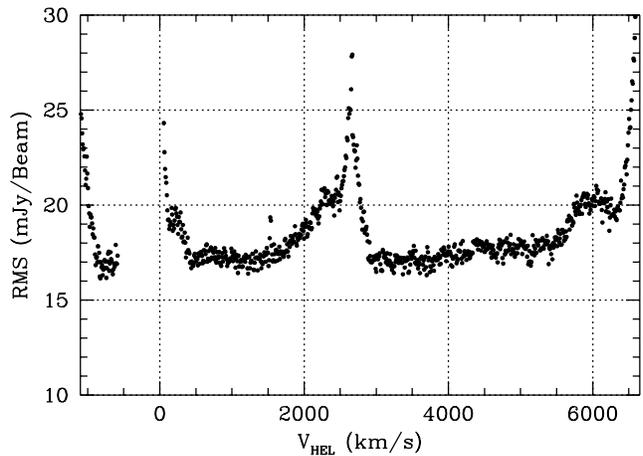}}
 \caption{Variation of {\sc rms} fluctuation level with heliocentric
 velocity over the survey velocity coverage.}
\label{fig:specrms}
\end{figure}

The resulting {\sc rms} fluctuation level as function of heliocentric
velocity is shown in Fig.~\ref{fig:specrms} as determined from a fit to
the peak of the histogram in each velocity channel. The effective
velocity coverage extends from about
$-1000~<~$V$_{Hel}~<$~6500~\kms. An increased noise level is seen for
$-200~<~$V$_{Hel}~<$150~\kms\ due to intermediate velocity emission
features associated with the Galaxy. Slightly elevated noise levels are
also seen for $1700~<~$V$_{Hel}~<$~2600~\kms\ and
$5500~<~$V$_{Hel}~<$~6500~\kms\ due to the problems of spectral
baseline stability in the upper third of each 20~MHz band as noted
above. A small velocity range near 1500~\kms\ also has a slightly
higher noise due to a significant degree of data flagging at this
velocity that was made necessary by recurring interference.

\begin{figure*}
\centering
\includegraphics[width=17cm]{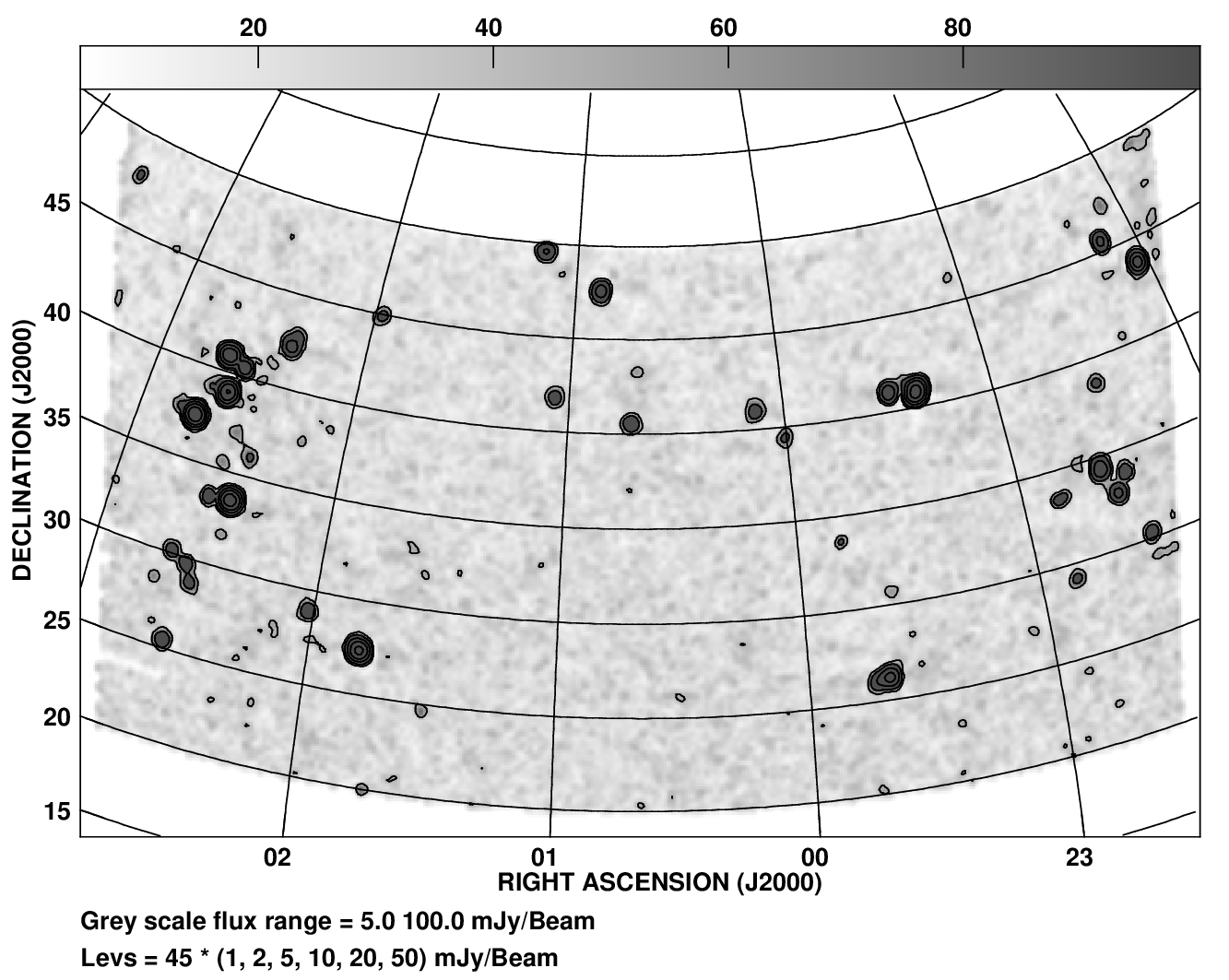}
 \caption{Illustration of the survey sky coverage and galaxy detections
between $250~<~$V$_{Hel}~<1500$~\kms. The peak brightness of \hi
emission at a velocity resolution of 42~\kms\  {\sc FWHM} is shown in
this velocity interval with a linear greyscale between and 5 and 100
mJy/Beam and contours drawn at (1, 2, 5, 10, 20 and 50) times 45
mJy/Beam. }
 \label{fig:lgsurc}
\end{figure*}

\section{Results}
\label{sec:results}

An overview of the survey sky coverage, together with an indication of
some of the galaxy detections, is given in Fig.~\ref{fig:lgsurc}. Peak
\hi brightness at a velocity resolution of 42~\kms\  {\sc FWHM} is shown
in the figure for the velocity interval
$250~<~$V$_{Hel}~<1500$~\kms. The first contour in the figure
corresponds to approximately 4$\sigma$ at this velocity resolution.
The total solid angle observed is 1800~deg$^2$. The {\sc rms}
fluctuation level varies slightly with recession velocity (as shown in
Fig.~\ref{fig:specrms}). The positions of bright continuum sources
(brighter than a few Jy) display residual fluctuations in excess of the
nominal noise level.  The resulting noise distribution is not entirely
Gaussian, and for this reason we consider it necessary to adopt a
conservative cut-off in our blind extraction of reliable \hi
detections. Rosenberg \& Schneider (\cite{rose02}) have shown from
their extensive simulations involving insertion of artificial sources
into surveys of this type that an asymptotic completeness level is
reached at a signal-to-noise ratio of about 8 (in terms of integrated
signal strength relative to the error in the integral), while below this
signal-to-noise ratio the completeness drops dramatically.

Candidate \hi detections in our combined data-cubes were determined by
two different methods. In the first instance, the SAD source finding
algorithm within Classic AIPS was used to extract all local peaks in
excess of 3 times the local {\sc RMS} level in datacubes having
Nyquist-sampled velocity resolutions of 2, 5, 10, 20 and 50 times the
basic velocity channel seperation of 8.4~\kms.  A reduced list of
emission candidates having a detected peak in excess of 5$\sigma$ in at
least two different velocity channels or two different velocity
smoothings was extracted for further analysis. The aim of this
procedure was to reliably recover significant detections spanning a
wide range in observed linewidth. A complimentary list of candidates
was determined from visual inspection of subsequent channel maps as
well as subsequent postion--velocity projections using the KVIEW
display program (Gooch \cite{gooc95}) for the data-cubes at velocity
resolutions of 2, 5, 10, 20 and 50 times the basic velocity channel
seperation of 8.4~\kms.  Candidate features were rejected if their
response in our data-cubes before spatial smoothing were inconsistent
with the telescope beam response in either $\alpha$ or $\delta$,
implying the time variable signature of interference.  The properties
of some 500 candidate detections were then estimated in detail. In
particular, the integrated line strength was determined for each
candidate by extracting the single spectrum from our spatially smoothed
data-cubes with the highest total flux density. The underlying
assumption is that essentially all galaxies will be unresolved with our
effective {\sc FWHM} beamsize of 3020$\times$2810 arcsec, corresponding
to 73$\times$68~kpc at the nearest galaxy distance of 5~Mpc. This
assumption was tested by comparing the peak with the integrated flux of
a Gaussian fit to an image of integrated \hi for each source. While
some sources show clear signs of confusion from nearby companions
(which will be discussed in detail below), there were no instances of
our having significantly resolved single galaxies with our survey beam.
The associated error in flux density was determined over a velocity
interval of 1.5$\times$W$_{20}$ (where W$_{20}$ is the velocity width
of the emission profile at 20\% of the peak intensity) together with
the actual {\sc rms} fluctuation level associated with this velocity
interval.

\begin{figure*}
\centering
 \includegraphics[width=16cm]{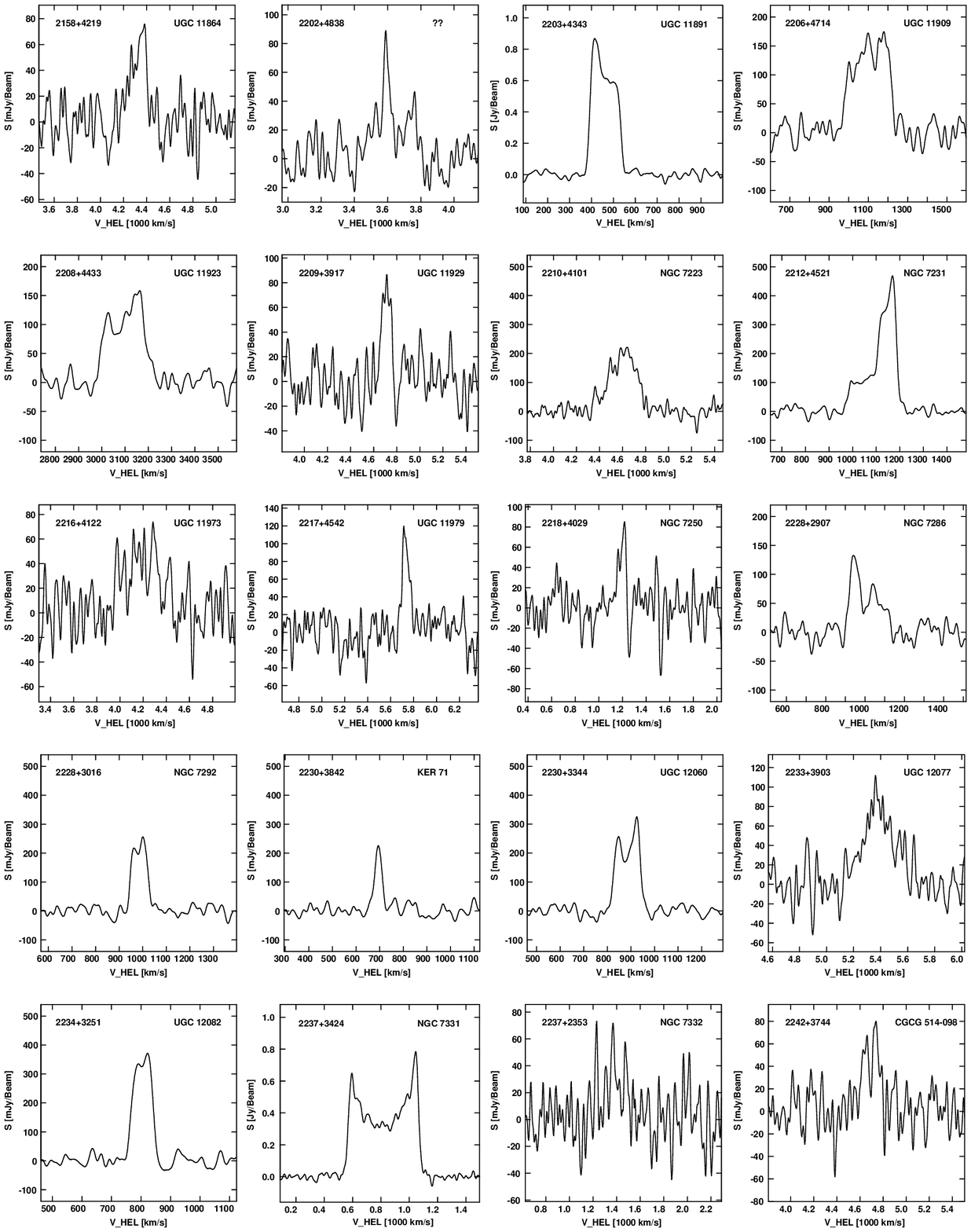}
 \caption{ \hi spectra of detected galaxies. The complete set of panels
for all detected galaxies is only available in the electronic version
of the paper at http://www.epdsciences.org. The catalog designation is
indicated in the upper left of each panel, while the likely optical ID
is indicated in the upper right.}
 \label{fig:lgcs}
\end{figure*}

Only the 155 candidates with an integrated flux density exceeding 8
times the associated error where retained. Spectra of each detection
are shown in Fig.~\ref{fig:lgcs} for the single spatial pixel with the
maximum integrated \hi signal. The source centroid positions were
determined from either a Gaussian or a parabolic fit to the peak in
images of integrated \hi line strength over the full velocity extent of
each detection. The positional accuracy is dependent on the
signal-to-noise ratio, and is expected to be roughly HWHM/(s/n),
implying about 3 arcmin {\sc rms} in both $\alpha$ and $\delta$ for the
lowest significance detections.

\begin{table*}
\renewcommand{\thefootnote}{\thempfootnote}
\caption{ Properties of \hi detected Galaxies 
}
\label{tab:higal}
\begin{tabular}{cccrrrrrlcrl}
\hline
{Name}                     & {R.A. (J2000)}
                         & {Dec. (J2000)}
                         & {V$_{Hel}$\ }
                         & {W$_{20}$\ }
                         & {S$_{int}$\ }
                         & {$\delta$S$_{int}$\ }
                         & {log(M$_{HI}$)}
                         & {Optical ID}
                         & {Note$^{\mathrm{a}}$}
                         & {Offset}
                         & {Comp.$^{\mathrm{b}}$}\\
{\ }                       & {$\rm(h\ m\ s)$}
                         & {$\rm(^\circ\ ^\prime\ ^{\prime\prime})$}
                         & \multispan2{\ \ (\kms)}
                         & \multispan2{\ \ (Jy \kms)} 
                         & {(M$_\odot$)\ \ \ }
                         & {\ }
                         & {\ }
                         & {(')}
                         & {\ } \\
(1)&(2)&(3)&(4)&(5)&(6)&(7)&(8)&(9)&(10)&(11)&(12)\\
\hline
 J2158+4219& 21 58 47&   42 19 46& 4335& 245&   9:& --&   9.92& UGC 11864   &\  &  9.8& 004 \\
 J2202+4838& 22 02 49&   48 38 16& 3605& 300&   9.3& 0.82&   9.77& ---??----&\     &  0.0& 000 \\ 
 J2203+4345& 22 03 36&   43 45 24&  455& 165&  99.5& 0.64&   9.37& UGC 11891&\     &  0.7& 008 \\
 J2206+4714& 22 06 23&   47 14 05& 1110& 240&  34.4& 0.77&   9.45& UGC 11909&\     &  1.5& 004 \\
 J2208+4433& 22 08 54&   44 33 50& 3110& 230&  24.3& 0.72&  10.07& UGC 11923&c     &  1.5& 108 \\
 J2209+3917& 22 09 22&   39 17 30& 4710& 125&   7.1& 0.55&   9.87& UGC 11929&c     &  3.0& 100 \\
 J2210+4101& 22 10 10&   41 01 03& 4600& 350&  59.6& 0.90&  10.78& NGC 7223&c      &  0.2& 326 \\
 J2212+4521& 22 12 38&   45 21 55& 1100& 200&  46.9& 0.67&   9.58& NGC 7231&c      &  2.6& 308 \\
 J2216+4122& 22 16 55&   41 22 33& 4190& 490&  17.9& 1.00&  10.18& UGC 11973&c     &  7.7& 102 \\
 J2217+4542& 22 17 26&   45 42 25& 5730& 120&   8.8& 0.55&  10.13& UGC 11979&\     &  7.6& 006 \\
 J2218+4029& 22 18 05&   40 29 17& 1198&  90&   5.4& 0.40&   8.70& NGC 7250 &\     &  5.1& 002 \\
 J2228+2907& 22 28 12&   29 07 10&  990& 220&  14.6& 0.72&   8.99& NGC 7286 &\     &  4.9& 005 \\
 J2228+3016& 22 28 22&   30 16 10&  985&  85&  18.5& 0.44&   9.09& NGC 7292 &\     &  1.6& 002 \\
 J2230+3842& 22 30 33&   38 42 50&  695&  55&   9.6& 0.33&   8.59& KKR 71   &\     &  1.5& 005 \\
 J2230+3344& 22 30 44&   33 44 55&  890& 120&  28.6& 0.52&   9.22& UGC 12060&\     &  1.9& 003 \\
 J2233+3903& 22 33 14&   39 03 31& 5385& 410&  21.7& 1.01&  10.47& UGC 12077&c     &  7.1& 135 \\
 J2234+3251& 22 34 13&   32 51 09&  805&  95&  26.8& 0.46&   9.12& UGC 12082&\     &  0.7& 000 \\
 J2237+3424& 22 37 01&   34 24 32&  835& 510& 230.5& 1.13&  10.08& NGC 7331 &\     &  0.8& 006 \\
 J2237+2353& 22 37 25&   23 53 23& 1378& 372&   8.9& 0.57&   9.00& NGC 7332&c      &  5.5& 204 \\
 J2242+3744& 22 42 19&   37 44 41& 4695& 275&  10.8& 0.82&  10.05& CGCG 514-098&\  &  6.1& 004 \\
 J2250+2909& 22 50 23&   29 09 55&  895& 120&  16.3& 0.52&   8.97& UGC 12212 &\    &  2.3& 000 \\
 J2304+2708& 23 04 17&   27 08 25& 1060& 150&   6.6& 0.53&   8.68& UGC 12340 &\    &  3.9& 009 \\
 J2313+2900& 23 13 32&   29 00 04& 3690& 220&  17.1& 0.70&  10.05& UGC 12430&c     &  2.5& 103 \\
 J2322+4050& 23 22 05&   40 50 03&  380& 255& 342.0& 0.81&   9.79& NGC 7640&o      &  0.7& 001 \\
 J2326+2504& 23 26 45&   25 04 20& 3490& 370&  21.3& 0.91&  10.10& NGC 7664 &\     &  1.3& 005 \\
 J2326+3046& 23 26 50&   30 46 06& 4520& 140&   5.3& 0.53&   9.70& UGC 12609&+     & 21.2& 001 \\
 J2327+2334& 23 27 35&   23 34 19& 3440& 240&  14.5& 0.73&   9.92& NGC 7673&c      &  1.8& 103 \\
 J2328+2234& 23 28 54&   22 34 46& 3475& 350&  10.8& 0.95&   9.80& NGC 7678 &\     & 11.2& 013 \\
 J2329+4101& 23 29 55&   41 01 20&  425& 135&  75.6& 0.55&   9.19& UGC 12632&o     &  2.0& 001 \\
 J2330+3008& 23 30 45&   30 08 45& 4530& 160&  11.7& 0.59&  10.05& UGC 12639&\     &  6.0& 002 \\
 J2331+2851& 23 31 31&   28 51 40& 5510& 240&   8.1& 0.70&  10.05& Mrk 0930&c      &  7.9& 104 \\
 J2333+2657& 23 33 20&   26 57 10& 3695& 145&   6.9& 0.56&   9.65& CGCG 476-066&\  &  5.2& 006 \\
 J2336+3216& 23 36 09&   32 16 26& 4955& 320&  12.1& 0.80&  10.14& UGC 12693&c     &  8.6& 101 \\
 J2337+3602& 23 37 32&   36 02 57& 5020& 700&  27.0& 1.20&  10.50& UGC 12697&++    & 19.6& 001 \\
 J2337+3048& 23 37 39&   30 48 40&  295& 165&   9.3& 0.66&   8.07& UGC 12713&\     &  9.8& 002 \\
 J2339+2509& 23 39 44&   25 09 31& 4935&  75&   3.6& 0.43&   9.61& CGCG 476-100&?  &  4.0& 008 \\
 J2340+2615& 23 40 43&   26 15 44&  750& 140&  89.1& 0.56&   9.56& UGC 12732&o     &  1.7& 009 \\
 J2346+3330& 23 46 15&   33 30 34& 4950& 370&  11.8& 0.96&  10.13& UGC 12776&\     &  8.4& 004 \\
 J2347+2932& 23 47 29&   29 32 59& 5115& 525&  19.1& 1.14&  10.36& NGC 7753&c      &  6.6& 100 \\
 J2348+2612& 23 48 58&   26 12 45&  800& 100&  18.8& 0.47&   8.93& UGC 12791&\     &  2.0& 004 \\
 J2349+4755& 23 49 12&   47 55 30& 4620& 165&   6.1& 0.64&   9.79& UGC 12796&\     &  1.3& 000 \\
 J2356+3203& 23 56 05&   32 03 29& 4850& 265&  11.2& 0.73&  10.09& UGC 12845&\     & 10.7& 009 \\
 J2358+4656& 23 58 58&   46 56 57& 5080& 430&  10.8& 1.06&  10.11& IC 1525&c       &  4.7& 303 \\
 J0000+3927& 00 00 22&   39 27 20&  330&  70&   7.5& 0.43&   8.03& UGC 12894 &\    &  2.4& 000 \\
 J0000+2320& 00 00 37&   23 20 27& 4560&  85&   5.4& 0.41&   9.71& UGC 12914&c     & 16.4& 102 \\
 J0003+3132& 00 03 27&   31 32 25& 4940& 390&  11.0& 0.98&  10.09& CGCG 498-067&c  &  8.5& 606 \\
 J0004+3129& 00 04 20&   31 29 32& 5020& 150&   8.8& 0.41&  10.01& NGC 7819&c      &  1.5& 304 \\
 J0006+4749& 00 06 55&   47 49 15& 4300&  85&  10.3& 0.44&   9.95& UGC 00048 &\    &  4.6& 002 \\
 J0007+2740& 00 07 27&   27 40 17& 4620& 190&  15.1& 0.62&  10.17& NGC 0001&c      &  3.3& 129 \\
 J0007+4056& 00 07 48&   40 56 23&  305&  90&  16.0& 0.47&   8.32& UGC 00064 &\    &  3.9& 002 \\
 J0010+2557& 00 10 07&   25 57 06& 4618& 360&  22.9& 0.97&  10.35& NGC 0023&c      &  3.4& 202 \\
 J0011+3319& 00 11 03&   33 19 27& 4806& 370&  18.6& 0.86&  10.30& NGC 0021&c      &  3.8& 409 \\
\hline
\end{tabular}
\begin{list}{}{}
\item[$^{\mathrm{a}}$] Explanation of notes: ``c'' for confused
sources, ``?'' for sources without a previous red-shift ,``o'' for
cases of a significant centroid offset of less than 10~kpc, ``+'' for
cases of centroid offset in excess of 10~kpc and $>5\sigma$, ``++'' for
centroid offset greater than 10~kpc and $>10\sigma$.
\item[$^{\mathrm{b}}$] A three digit ``confusion''
index, ``abc'' enumerating the number of cataloged campanions
(truncated at 9) within a
30$^\prime$ radius which are (a) within 400~\kms, (b) between 400 and
1000~\kms and (c) of unknown red-shift.
\end{list}
\end{table*}

\begin{table*}
\caption{ Properties of \hi detected Galaxies (continued.)
}
\begin{tabular}{cccrrrrrlcrl}
\hline
{Name}                     & {R.A. (J2000)}
                         & {Dec. (J2000)}
                         & {V$_{Hel}$\ }
                         & {W$_{20}$\ }
                         & {S$_{int}$\ }
                         & {$\delta$S$_{int}$\ }
                         & {log(M$_{HI}$)}
                         & {Optical ID}
                         & {Note$^{\mathrm{a}}$}
                         & {Offset}
                         & {Comp.$^{\mathrm{b}}$}\\
{\ }                       & {$\rm(h\ m\ s)$}
                         & {$\rm(^\circ\ ^\prime\ ^{\prime\prime})$}
                         & \multispan2{\ \ (\kms)}
                         & \multispan2{\ \ (Jy \kms)} 
                         & {(M$_\odot$)\ \ \ }
                         & {\ }
                         & {\ }
                         & {(')}
                         & {\ }\\
(1)&(2)&(3)&(4)&(5)&(6)&(7)&(8)&(9)&(10)&(11)&(12)\\
\hline
 J0012+4147& 00 12 16&   41 47 48& 5000& 100&   5.0& 0.45&   9.76& UGC 00112 &\    &  2.9& 001 \\
 J0013+2655& 00 13 51&   26 55 49& 4680& 244&  12.2& 0.78&  10.09& UGC 00127 &\    &  2.5& 001 \\
 J0013+3606& 00 13 58&   36 06 18& 4580& 120&   7.0& 0.49&   9.83& UGC 00128 &\    &  6.8& 003 \\
 J0028+4320& 00 28 59&   43 20 55&  190&  40&   4.1& 0.34&   7.52& UGC 00288 &\    &  5.0& 003 \\
 J0042+4031& 00 42 31&   40 31 36&  230& 100&  22.4& 0.50&   8.31& And IV    &\    &  2.7& 019 \\
 J0043+2705& 00 43 58&   27 05 05& 5190& 240&  14.8& 0.79&  10.26& UGC 00470&c     & 14.7& 102 \\
 J0047+2951& 00 47 52&   29 51 32& 5030& 105&   3.9& 0.46&   9.65& CGCG 501-016&c  &  6.1& 706 \\
 J0048+3159& 00 48 47&   31 59 16& 4530& 105&  21.5& 0.48&  10.31& NGC 0262&c      &  1.8& 223 \\
 J0052+4733& 00 52 14&   47 33 34&  640& 150&  47.9& 0.57&   9.19& NGC 0278 &\     &  1.7& 003 \\
 J0100+4757& 01 00 13&   47 57 15& 2740& 200&  23:& --&   9.93& UGC 00622&c     &  3.5& 302 \\
 J0100+4746& 01 00 54&   47 46 20& 2650& 410&  90:& --&  10.50& IC 0065&c       &  5.4& 302 \\
 J0103+4149& 01 03 56&   41 49 58&  840& 135&  23.6& 0.56&   9.05& UGC 00655 &\    &  1.1& 002 \\
 J0110+4316& 01 10 33&   43 16 14& 4945& 390&  14.9& 1.01&  10.22& UGC 00728&c     &  1.2& 104 \\
 J0110+4934& 01 10 34&   49 34 12&  645& 150&  39.8& 0.57&   9.11& UGC 00731&o     &  2.5& 003 \\
 J0114+2710& 01 14 56&   27 10 24& 3650& 390&  12.9& 0.86&   9.90& ADBS J0114 &\   &  3.3& 005 \\
 J0116+3727& 01 16 58&   37 27 02& 4830& 120&   7.3& 0.49&   9.89& UGCA 016   &\   &  8.9& 005 \\
 J0120+3327& 01 20 42&   33 27 10& 5420&  50&   5.5& 0.32&   9.86& CGCG 502-039&c  &  2.8& 953 \\
 J0125+3400& 01 25 48&   34 00 19& 4820& 640&  18.7& 1.26&  10.30& NGC 0523&c      &  5.8& 514 \\
 J0127+3126& 01 27 31&   31 26 47& 4108& 494&  14.3& 1.13&  10.04& UGC 01033 &\    &  6.6& 004 \\
 J0130+4100& 01 30 02&   41 00 14& 2810& 170&  17:& --&   9.81& UGC 01070&c     &  1.8& 102 \\
 J0130+2551& 01 30 04&   25 51 30& 3660& 155&   8.4& 0.56&   9.71& UGC 01073 &\    &  0.9& 000 \\
 J0130+2402& 01 30 50&   24 02 25& 3415& 115&   8.1& 0.52&   9.64& UGC 01084 &\    &  9.0& 001 \\
 J0130+3404& 01 30 59&   34 04 46& 5035& 400&  14.1& 1.00&  10.21& CGCG 521-039&c  &  1.7& 109 \\
 J0135+4752& 01 35 51&   47 52 48& 5310& 500&  14.1& 1.14&  10.26& UGC 01132&+     & 20.1& 002 \\
 J0136+4759& 01 36 18&   47 59 50& 1700& 120&  11.2& 0.56&   9.24& Anon&?          &  6.0& 001 \\
 J0143+1959& 01 43 15&   19 59 00&  490&  80&   5.:& --&   7.96& UGCA 020 &\     &  0.5& 002 \\
 J0143+2843& 01 43 32&   28 43 12& 4030& 205&  10.8& 0.64&   9.90& NGC 0661&c      &  9.4& 209 \\
 J0143+2736& 01 43 35&   27 36 01& 4025& 240&   6.3& 0.70&   9.67& FGC 0191&+      & 23.0& 009 \\
 J0145+2533& 01 45 43&   25 33 50& 3830& 120&  10.6& 0.52&   9.85& UGC 01230 &\    &  3.5& 011 \\
 J0147+2723& 01 47 44&   27 23 39&  390& 290& 257.0& 0.87&   9.52& VV 338 Gpair&o  &  0.8& 004 \\
 J0149+3234& 01 49 38&   32 34 55&  155& 140&  36.9& 0.66&   8.24& UGC 01281 &\    &  1.5& 009 \\
 J0150+3515& 01 50 32&   35 15 10& 4200& 460&  10.3& 1.06&   9.92& NGC 0688&c      &  3.1& 659 \\
 J0150+2159& 01 50 52&   21 59 55& 2935& 175&  24.2& 0.63&   9.98& NGC 0694&c      &  1.5& 701 \\
 J0151+2216& 01 51 13&   22 16 20& 3100& 480&  55.7& 1.04&  10.39& NGC 0697&c      &  5.3& 402 \\
 J0154+3720& 01 54 20&   37 20 36& 5515& 205&   7.2& 0.66&   9.99& UGC 01398&c     & 23.4& 988 \\
 J0154+2049& 01 54 25&   20 49 04& 4930& 220&   6.1& 0.76&   9.82& NGC 0722 &\     &  8.8& 003 \\
 J0154+2310& 01 54 27&   23 10 44& 4975& 185&   9.1& 0.70&  10.00& [ZBS97] A31&c   &  9.9& 201 \\
 J0157+3557& 01 57 41&   35 57 12& 4908& 350&  15.4& 0.57&   9.61& NGC 0753&c      &  2.2& 539 \\
 J0157+4454& 01 57 59&   44 54 10&  705& 135&  19.8& 0.54&   8.83& NGC 0746 &\     &  1.7& 002 \\
 J0158+2453& 01 58 49&   24 53 57& 5110& 175&  39.7& 0.66&  10.66& NGC 0765&c      &  0.5& 302 \\
 J0200+2814& 02 00 53&   28 14 26& 5300& 185&   5.3& 0.61&   9.82& NGC 0780&c      &  3.9& 105 \\
 J0200+3155& 02 00 56&   31 55 17& 5200& 210&  12.6& 0.74&  10.18& NGC 0783&c      &  3.2& 508 \\
 J0201+2850& 02 01 18&   28 50 05&  190& 120&  63.7& 0.58&   8.52& NGC 0784&c      &  0.3& 105 \\
 J0203+2205& 02 03 01&   22 05 42& 2680& 350&  55:& --&  10.26& UGC 01547 &\    &  5.5& 001 \\
 J0203+2402& 02 03 38&   24 02 40& 2690&  70&   7:& --&   9.37& UGC 01551&c     &  1.9& 103 \\
 J0205+2441& 02 05 44&   24 41 28& 4825& 255&  11.9& 0.72&  10.09& UGC 01575&c     &  7.8& 102 \\
 J0205+3457& 02 05 53&   34 57 47& 4385& 410&  12.4& 0.91&  10.03& UGC 01581&c     &  6.3& 109 \\
 J0206+4435& 02 06 40&   44 35 40& 5200& 525&  32.8& 1.17&  10.60& NGC 0812&c      &  2.4& 107 \\
 J0208+3203& 02 08 49&   32 03 24& 5010& 115&  12.2& 0.55&  10.14& UGC 01641 &\    &  5.9& 002 \\
 J0213+4156& 02 13 49&   41 56 44& 4355& 120&   5.1& 0.49&   9.64& CGCG 538-034 &\ &  4.6& 005 \\
 J0213+3724& 02 13 51&   37 24 16& 4635& 210&  17.6& 0.65&  10.23& UGC 01721&c     &  8.6& 106 \\
 J0215+2511& 02 15 32&   25 11 29& 5005& 350&  12.7& 0.95&  10.15& UGC 01739 &\    &  2.9& 002 \\

\hline
\end{tabular}
\begin{list}{}{}
\item[$^{\mathrm{a}}$] Explanation of notes: ``c'' for confused
sources, ``?'' for sources without a previous red-shift ,``o'' for
cases of a significant centroid offset of less than 10~kpc, ``+'' for
cases of centroid offset in excess of 10~kpc and $>5\sigma$, ``++'' for
centroid offset greater than 10~kpc and $>10\sigma$.
\item[$^{\mathrm{b}}$] A three digit ``confusion''
index, ``abc'' enumerating the number of cataloged campanions
(truncated at 9) within a
30$^\prime$ radius which are (a) within 400~\kms, (b) between 400 and
1000~\kms and (c) of unknown red-shift.
\end{list}
\end{table*}

\begin{table*}
\caption{ Properties of \hi detected Galaxies (continued.)
}
\begin{tabular}{cccrrrrrlcrl}
\hline
{Name}                     & {R.A. (J2000)}
                         & {Dec. (J2000)}
                         & {V$_{Hel}$\ }
                         & {W$_{20}$\ }
                         & {S$_{int}$\ }
                         & {$\delta$S$_{int}$\ }
                         & {log(M$_{HI}$)}
                         & {Optical ID}
                         & {Note$^{\mathrm{a}}$}
                         & {Offset}
                         & {Comp.$^{\mathrm{b}}$}\\
{\ }                       & {$\rm(h\ m\ s)$}
                         & {$\rm(^\circ\ ^\prime\ ^{\prime\prime})$}
                         & \multispan2{\ \ (\kms)}
                         & \multispan2{\ \ (Jy \kms)} 
                         & {(M$_\odot$)\ \ \ }
                         & {\ }
                         & {\ }
                         & {(')}
                         & {\ }\\
(1)&(2)&(3)&(4)&(5)&(6)&(7)&(8)&(9)&(10)&(11)&(12)\\
\hline
 J0215+2834& 02 15 57&   28 34 23& 3023& 255&  11.9& 0.74&   9.70& NGC 0865&c      &  4.2& 103 \\
 J0217+2938& 02 17 35&   29 38 05& 5250&  70&   5.0& 0.42&   9.79& MRK 1030 &\     &  6.8& 009 \\
 J0221+2828& 02 21 03&   28 28 44& 4750& 205&  17.8& 0.64&  10.25& UGC 01791&++    & 20.6& 003 \\
 J0221+4245& 02 21 13&   42 45 45&  625&  70&   7.0& 0.39&   8.28& UGC 01807&c     &  0.0& 109 \\
 J0222+2515& 02 22 28&   25 15 34& 4625& 370&  12.4& 0.93&  10.07& CGCG 483-018&c  &  6.9& 117 \\
 J0222+4754& 02 22 38&   47 54 15& 5120& 165&  11.2& 0.65&  10.12& UGC 01830&c     &  3.4& 104 \\
 J0222+4219& 02 22 35&   42 19 40&  560& 480& 194.9& 0.98&   9.65& NGC 0891&c      &  1.3& 109 \\
 J0224+3559& 02 24 52&   35 59 25&  575& 100&  14.9& 0.77&   8.53& UGC 01865 &\    &  3.3& 001 \\
 J0225+3136& 02 25 10&   31 36 40& 4800& 290&  12.7& 0.83&  10.12& UGC 01856 &\    &  8.2& 013 \\
 J0227+3334& 02 27 12&   33 34 51&  555& 220& 302.5& 0.69&   9.81& NGC 0925&o      &  0.4& 003 \\
 J0227+4159& 02 27 36&   41 59 31& 5650& 275&  19.9& 0.83&  10.45& NGC 0923&c      &  0.9& 749 \\
 J0227+3142& 02 27 41&   31 42 25&  600& 140&  10.0& 0.54&   8.37& UGC 01924 &\    &  2.2& 003 \\
 J0228+3115& 02 28 43&   31 15 00& 5030& 475&  25.7& 1.14&  10.46& NGC 0931&c      &  7.1& 302 \\
 J0228+4556& 02 28 51&   45 56 37& 5090& 480&  13.0& 1.12&  10.18& IC 1799&c       &  1.8& 106 \\
 J0230+3705& 02 30 49&   37 05 08&  625& 220&  15.6& 0.69&   8.61& NGC 0949 &\     &  3.1& 003 \\
 J0231+2835& 02 31 12&   28 35 08& 4620& 290&   9.7& 0.77&   9.96& UGC 01971&c     & 16.1& 209 \\
 J0232+3526& 02 32 04&   35 26 35&  570& 100&   9.6& 0.46&   8.33& NGC 0959 &\     &  5.2& 003 \\
 J0232+2328& 02 32 26&   23 28 01& 5560& 500&  23.6& 1.36&  10.50& UGC 02020&c     & 10.6& 101 \\
 J0232+2852& 02 32 42&   28 52 13& 1015& 105&  15.8& 0.48&   8.94& UGC 02017 &\    &  1.9& 009 \\
 J0232+3845& 02 32 56&   38 45 20&  575&  90&   4.6& 0.44&   8.03& UGC 02014 &\    &  4.6& 001 \\
 J0233+3330& 02 33 20&   33 30 00&  605&  55&  18.4& 0.33&   8.64& UGC 02023 &\    &  0.7& 002 \\
 J0233+4032& 02 33 51&   40 32 10&  580&  70&  36.0& 0.39&   8.93& UGC 02034 &\    &  1.5& 004 \\
 J0233+3210& 02 33 25&   32 10 07& 4670& 220&   7.5& 0.67&   9.86& IC 1815&c       & 19.5& 305 \\
 J0234+2923& 02 34 02&   29 23 03& 1530& 375&  17.3& 1.16&   9.30& NGC 0972 &\     &  5.0& 019 \\
 J0234+2056& 02 34 21&   20 56 39& 4355& 345&  14.7& 0.90&  10.09& NGC 0976&c      &  5.3& 202 \\
 J0234+2943& 02 34 40&   29 43 45& 1025& 100&  19.8& 0.48&   9.05& UGC 02053 &\    &  2.6& 019 \\
 J0235+3727& 02 35 32&   37 27 39& 3850& 220&  14.4& 0.71&   9.99& UGC 02065&c     &  2.4& 303 \\
 J0236+2523& 02 36 20&   25 23 47&  705& 215&  48.7& 0.68&   9.15& UGC 02082 &\    &  1.9& 009 \\
 J0236+3857& 02 36 25&   38 57 53&  905& 140& 140.9& 0.56&   9.83& IC 0239  &\     &  0.6& 000 \\
 J0236+2104& 02 36 48&   21 04 24& 4150& 325&  24.6& 0.90&  10.27& NGC 0992&+      &  8.9& 001 \\
 J0239+4052& 02 39 14&   40 52 00&  625& 230& 186.0& 0.70&   9.69& NGC 1003&c      &  0.6& 107 \\
 J0239+3009& 02 39 14&   30 09 55&  970& 195&  39.8& 0.67&   9.31& NGC 1012 &\     &  0.9& 019 \\
 J0239+3015& 02 39 55&   30 15 43&  810&  40&   5.4& 0.30&   8.31& [VR94] 0236&c ? &  0.9& 108 \\
 J0239+3905& 02 39 56&   39 05 25&  605& 370&  40.0& 0.86&   8.99& NGC 1023&c      &  5.7& 405 \\
 J0240+4221& 02 40 18&   42 21 41& 4175& 180&   6.8& 0.67&   9.73& IRAS 0237&c     & 15.0& 115 \\
 J0240+3920& 02 40 33&   39 20 20&  925& 120&  14.1& 0.52&   8.84& NGC 1023C&c     &  2.8& 400 \\
 J0241+3213& 02 41 08&   32 13 55& 4481& 240&  14.7& 0.70&  10.12& CGCG 505-033&c  &  3.4& 212 \\
 J0242+4327& 02 42 16&   43 27 50&  565&  60&   3.2& 0.36&   7.86& UGC 02172 &\    &  6.6& 001 \\
 J0242+2829& 02 42 53&   28 29 40& 1545& 310&  31.9& 1.03&   9.57& NGC 1056 &\     &  4.9& 009 \\
 J0243+3720& 02 43 30&   37 20 35&  520&  45&  94.4& 0.31&   9.26& NGC 1058 &\     &  0.2& 009 \\
 J0244+3208& 02 44 56&   32 08 30& 1580& 140&  13.3& 0.60&   9.22& kkh 014  &\     &  1.9& 007 \\
 J0247+4114& 02 47 50&   41 14 40& 4045& 290&  14.3& 0.85&  10.03& NGC 1086&c      &  1.2& 240 \\
 J0247+3736& 02 47 44&   37 36 29&  560& 150&  29.0& 0.55&   8.79& UGC 02259&c     &  4.7& 107 \\
 J0254+4238& 02 54 13&   42 38 30& 2150& 200&  11.5& 0.99&   9.41& UGC 02370 &\    &  2.5& 001 \\
 J0259+4452& 02 59 58&   44 52 55& 1800& 100&  11.1& 0.58&   9.25& NGC 1161&c      & 13.5& 118 \\
 J0302+4852& 03 02 15&   48 52 36& 2440& 270&  16.8& 0.86&   9.68& HFLLZOAG144&c ? &  2.8& 107 \\
 J0302+4232& 03 02 10&   42 32 39& 4170& 275&   8.8& 0.83&   9.84& NGC 1164&c      &  3.1& 132 \\
 J0304+4313& 03 04 58&   43 13 39& 2760& 265&  19:& --&   9.83& NGC 1171 &\     & 14.8& 003 \\
 J0305+4215& 03 05 48&   42 15 10& 2810& 240&  14:& --&   9.71& IC 0284&c       &  8.2& 107 \\
 J0309+3840& 03 09 31&   38 40 20& 3410& 150&  16.9& 0.59&   9.95& NGC 1213 &\     &  3.0& 001 \\
 J0332+4747& 03 32 00&   47 47 55&  220&  90&  23.3& 0.46&   8.14& UGC 02773 &\    &  1.3& 009 \\

\hline
\end{tabular}
\begin{list}{}{}
\item[$^{\mathrm{a}}$] Explanation of notes: ``c'' for confused
sources, ``?'' for sources without a previous red-shift ,``o'' for
cases of a significant centroid offset of less than 10~kpc, ``+'' for
cases of centroid offset in excess of 10~kpc and $>5\sigma$, ``++'' for
centroid offset greater than 10~kpc and $>10\sigma$.
\item[$^{\mathrm{b}}$] A three digit ``confusion''
index, ``abc'' enumerating the number of cataloged campanions
(truncated at 9) within a
30$^\prime$ radius which are (a) within 400~\kms, (b) between 400 and
1000~\kms and (c) of unknown red-shift.
\end{list}
\end{table*}

\subsection{Galaxy properties}

The properties of our \hi selected detections are summarized in
Table~\ref{tab:higal}. In addition to the position of the \hi centroid
and integrated flux density, $F_{HI}$ in units of Jy-\kms, we tabulate
the heliocentric recession velocity, V$_{Hel}$, and velocity width at
20\% of the peak brightness, W$_{20}$. The integrated flux density has
been converted to an \hi mass by first calculating the recession
velocity in the Local Group Standard of Rest frame, V$_{LGSR}$ =
V$_{Hel} + 300$ sin $l$ cos $b$, then assuming a Hubble constant,
$H_0$~=~75~\kms\ Mpc$^{-1}$ to derive the distance, $D~=~$V$_{LGSR}/H_O$
in Mpc, and finally using M$_{HI}$~=~2.356$\times10^5 D^2 F_{HI}$ under
the simplifying assumption of negligible \hi opacity.

Seven of our detections have recession velocities that lie near the
transition in our velocity coverage from the lower to the upper band of
20~MHz width, at V$_{Hel}~\sim~2800$~\kms. While our velocity coverage
is complete across this transition, the spectral baseline extent is
severely impaired for such objects. Consequently, there is a large
systematic uncertainty in the integrated flux density, recession
velocity and velocity width of these detections. Two of our detections
(UGC~11864 and UGCA~020) lie at the edge of our nominal sky coverage, so
that they are not properly sampled and also have large uncertainties in
their properties.  We indicate these large uncertainties by a ``:''
suffix in Table~\ref{tab:higal} and do not use our measured properties
for these sources in the subsequent analysis.

\begin{figure*}
\centering
 \begin{tabular}{cc}
 \resizebox{85mm}{!}{\includegraphics{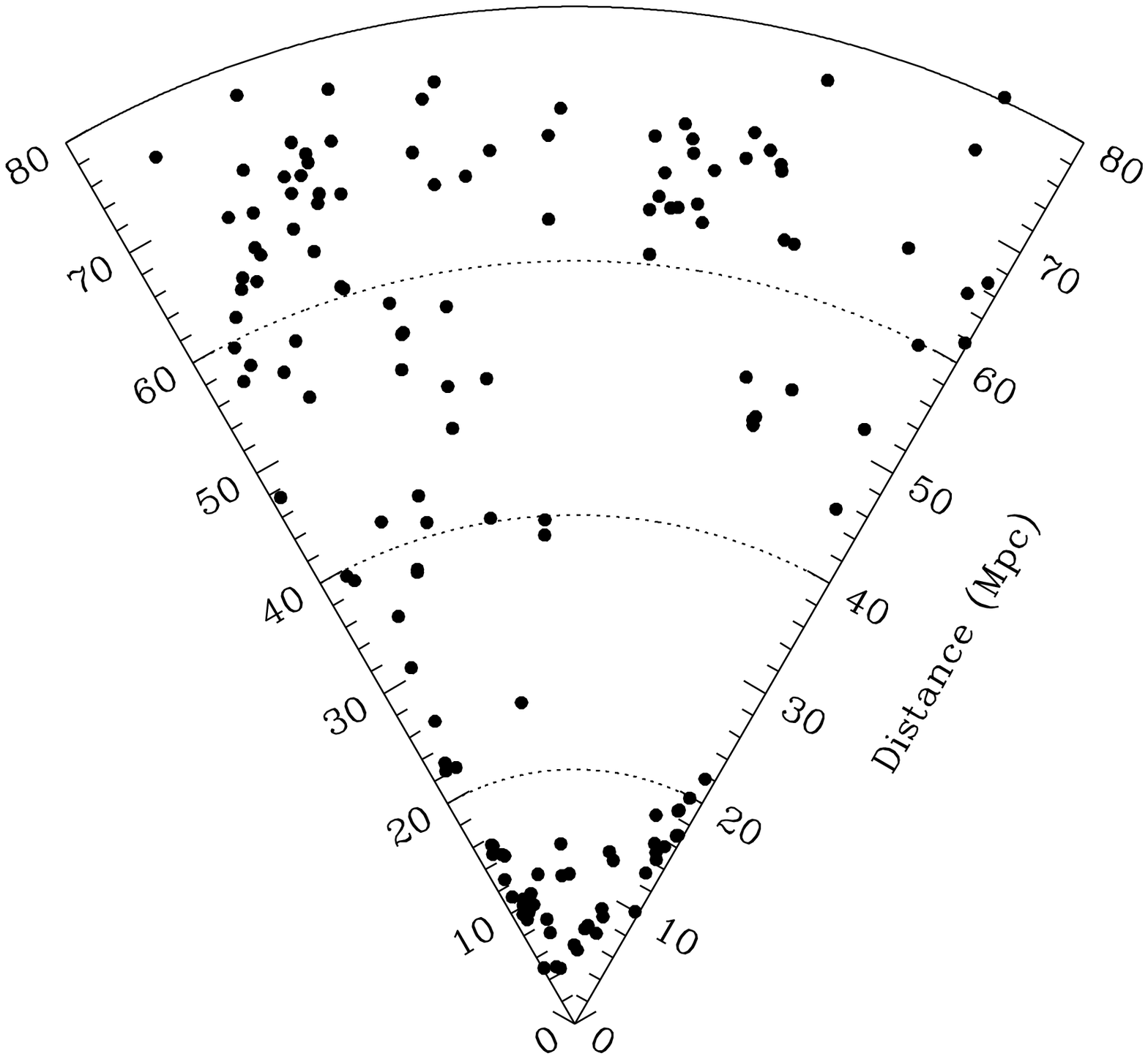}} &
 \resizebox{85mm}{!}{\includegraphics{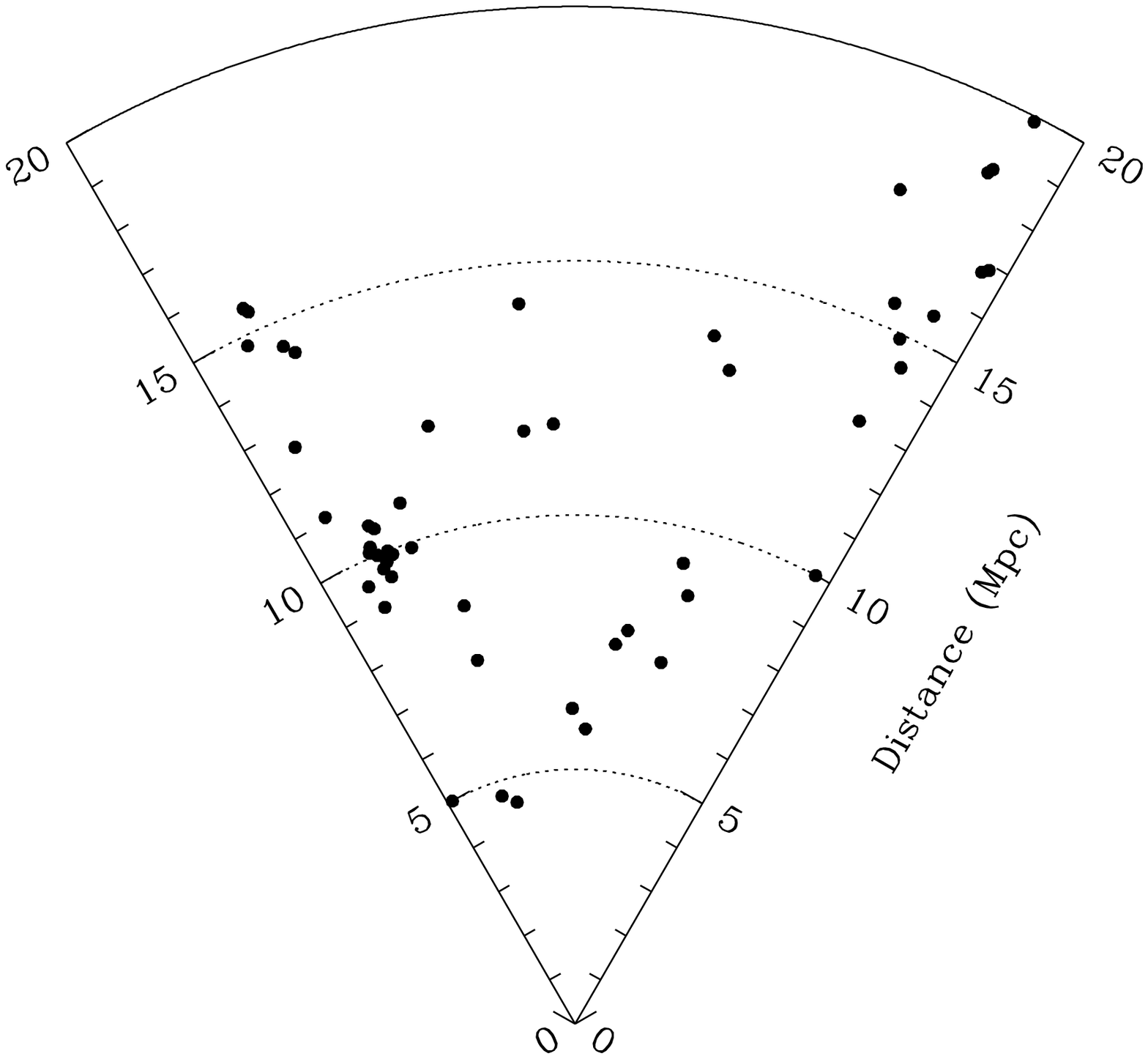}} \\
 \end{tabular}
 \caption{ Wedge diagrams for the \hi detected galaxies in our
survey. All detected galaxies are depicted on the left, while only
those within 20~Mpc are shown on the right.}
 \label{fig:wedge}
\end{figure*}

The spatial distribution of our \hi detections derived from the
distance calculated as above is shown in Fig.~\ref{fig:wedge}, both for
the entire depth of our survey (about 80~Mpc) and out to 20~Mpc. A
moderate galaxy concentration out to about 15~Mpc is followed by an
apparent void over much of our surveyed field out to about 45~Mpc,
which in turn is followed by an substantial increase in detected number
density out to about 80~Mpc. A galaxy filament along the eastern edge
of our coverage connects the nearby and more distant concentrations.

\subsection{Optical ID's}

Cataloged counterparts of our \hi detections were sought in the
NASA/IPAC Extragalactic Database (NED) on 2003/03/01 within an extended
error circle of 30~arcmin radius. This search radius was chosen since
it corresponds to the radius of the first null in the primary beam of
the WSRT telescopes. Only objects within this radius can contribute
significantly to our detected \hi fluxes. Identifications with
cataloged galaxies having published red-shifts was possible in most
cases. The NED ID's of our \hi detections are listed in
Table~\ref{tab:higal} together with the angular offset of the NED
position from that of the \hi centroid. Some objects deserving special
attention are noted below.

J2202+4838, corresponding to $(l,b)~=~(96.5,-5.4)$, has no cataloged
optical counterpart nor candidate galaxy visible in the DSS. Given the
low galactic latitude of this line-of-sight this is perhaps not too
surprising.

J2339+2509 appears to be associated with CGCG~476-100, although no
previous red-shift is available for this galaxy.

J0136+4759 appears to be associated with an uncataloged LSB galaxy at
($\alpha,\delta$)~=~(01:36:40,+48:03:40), lying very near a bright
foreground star.

J0239+3015 is very likely associated with the NED galaxy
[VR94] 0236.9+3003 with tabulated photometry by Vennik \& Richter
(\cite{venn94}), but without a previous red-shift determination.

J0302+4852 appears to be associated with the cataloged source HFLLZOA
G144.00-08.53 which has no previous red-shift determination.

In the final column of Table~\ref{tab:higal} we give an indication of
known and possible companions of our detections. We list the number
(truncated at a maximum value of nine) of NED galaxies within a 30
arcmin search radius of the primary optical ID which have (a) a known
red-shift within 400~\kms\ of the primary ID, (b) a known red-shift
between 400 and 1000~\kms\ of the primary ID, and (c) an unknown
reshift. These three categories of possible companion galaxies have
been used to define a confusion index relevant to our survey made up of
the three digits ``abc''.  All sources with a confusion index of 100 or
greater are indicated by a ``c'' entry in the Note column of
Table~\ref{tab:higal}.  A total of 85 of our 155 detections are
classified as ``unconfused'' by this criterion.  The number of possible
companions in the third category considered (no known red-shift)
deserves some further comment. Although some of our detected galaxies
have a large number of objects (as many as 66) in this category, it is
often merely an indication that the general field has received
intensive study, usually directed at a distant background galaxy
cluster. While this category remains an ambiguous and non-uniform
measure of possible companions, it still gives some indication of what
is known of the galaxy environment.

\begin{figure*}
\centering
 \begin{tabular}{cc}
 \resizebox{85mm}{!}{\includegraphics{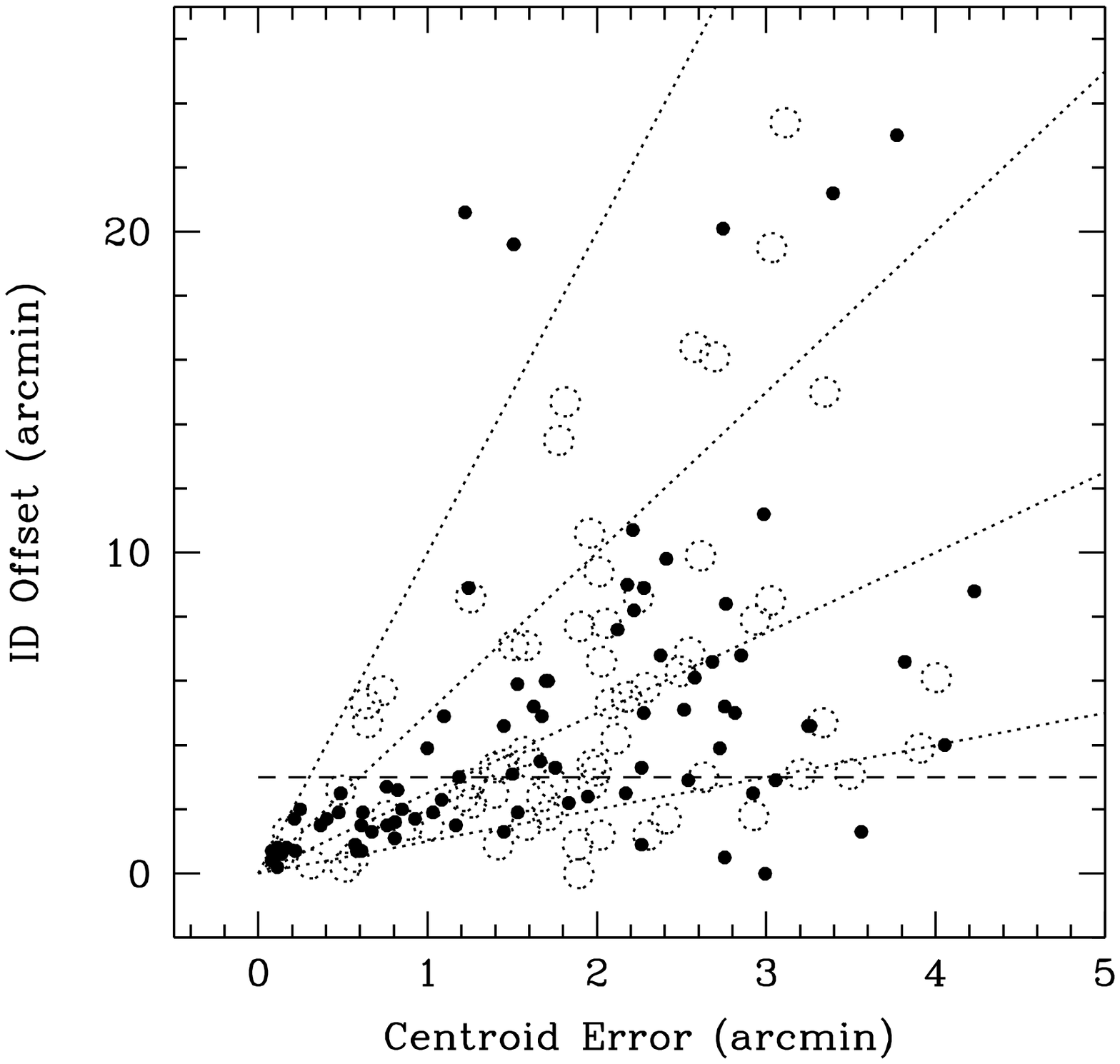}} &
 \resizebox{85mm}{!}{\includegraphics{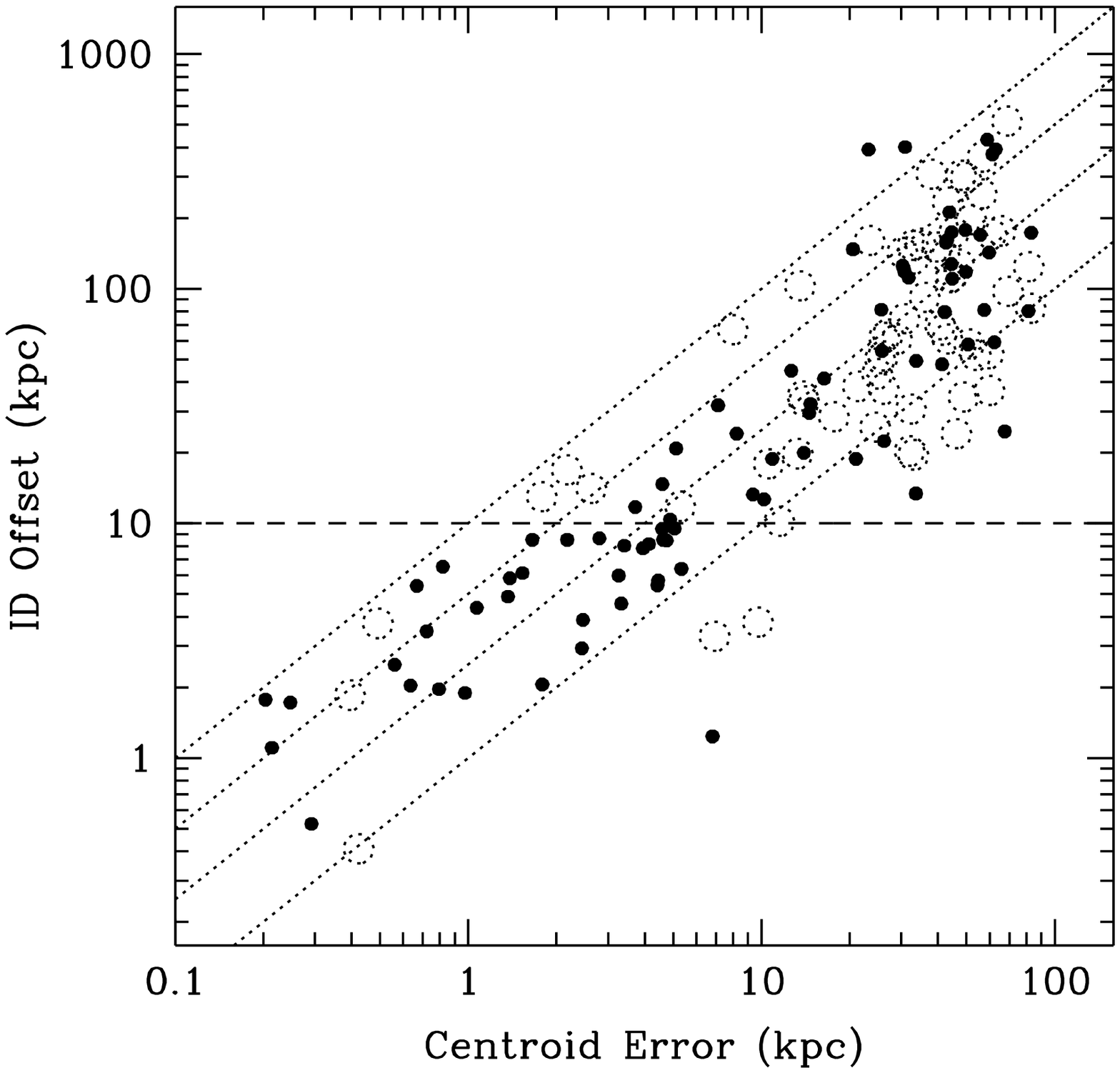}} \\
 \end{tabular}
 \caption{Comparison of the expected {\sc rms } positional error with
the observed positional offsets of our \hi detections from the primary
optical counterpart. The dotted lines correspond 1, 2.5, 5 and
10$\sigma$. Dashed lines are drawn at an offset of 3~arcmin and 10~kpc,
which may distinguish cases of internal asymmetries from confusion with
gas-rich companions. Galaxies with no cataloged companions within
30~arcmin and 400~\kms\  are plotted as filled circles, while those
with such companions are plotted as open circles.}
 \label{fig:offset} 
\end{figure*}

We compare the anticipated centroid error of our \hi detections (from
$\sqrt 2$ HWHM/(s/n)) with the angular and linear offsets of the NED
ID's in Fig.~\ref{fig:offset}. The dotted lines in the figure
correspond to 1, 2.5, 5 and 10 times the estimated centroid error.
Nominally unconfused galaxies (those with no known companions within
30~arcmin and 400~\kms) are plotted as filled circles in the figure,
while galaxies with known nearby companions are plotted as the open
circles.  The mean observed offset for the unconfused galaxies is 4.8
arcmin and 66~kpc, although the majority of these have a low
significance. A small concentration of significant centroid offsets is
seen below about 3~arcmin and 10~kpc. These cases are indicated by
an ``o'' entry in the Note column of Table~\ref{tab:higal}. This component
may be due to asymmetries in the \hi distribution of individual
objects, since it corresponds to sub-galactic dimensions.  Most of the
large observed offsets occur in cases of galaxies with cataloged
companions.  In addition, there are a small number of instances of
larger angular offsets of high significance in apparently unconfused
galaxies.  Positional offsets larger than 10~kpc and 5$\sigma$ are seen
in 6 instances, and larger than 10~kpc and 10$\sigma$ in two. These
cases have been indicated in the Note column of Table~\ref{tab:higal} by
a ``+'' symbol entry for offsets larger than 10~kpc and
5$\sigma$ and a pair of ``+'' symbols for offsets larger than 10~kpc
and 10$\sigma$.  This component of offsets is suggestive of either
severe asymmetries in the gas distribution of single galaxies or nearby
uncataloged gas-rich companions within the telescope beam.

\begin{figure*}
\centering
  \caption{ Atlas of ambiguous optical ID's. All \hi detections are
depicted which have a significant angular offset (more than 5$\sigma$)
from the optical ID, or which have no prior red-shift determination.
Countours at 80, 90 and 97\% of the peak in integrated \hi are overlaid
on a 30$\times$30 arcmin red frame from the second generation DSS. A
square-root transfer function is used for the optical image.(This
figure follows the preprint as a .png file, rather than embedded PS.}
  \label{fig:lgco}
\end{figure*}

In Fig.~\ref{fig:lgco} we present an atlas of images taken from the
second generation digital sky survey of the Space Telescope Science
Institute for all of the \hi detections with somewhat ambiguous optical
ID's. Those fields are depicted which have a significant angular offset
(more than 5$\sigma$) from the optical ID, or which have no prior
red-shift determination.  In each case a red 30$\times$30 arcmin field
centered approximately on the \hi centroid was extracted.

\begin{figure}
 \resizebox{\hsize}{!}{\includegraphics{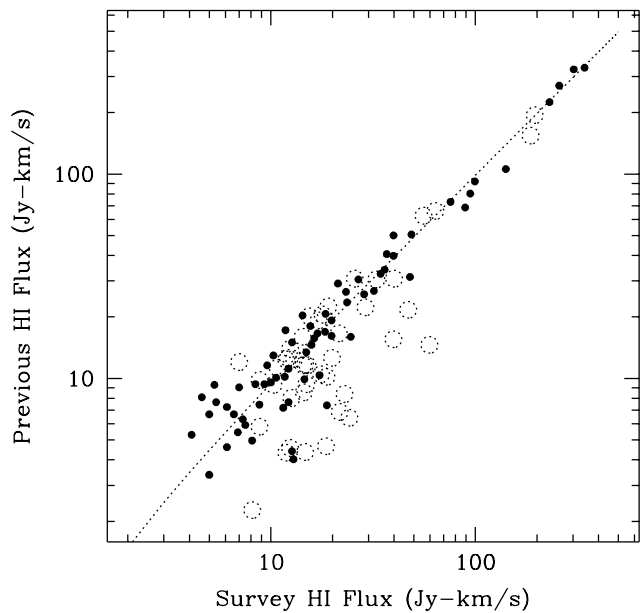}}
 \caption{Comparison of our survey integrated \hi fluxes with previous
determinations in the literature. The objects indicated by open circles
have cataloged companions within 30~arcmin and 400~\kms\ and are
classified as confused. The linear regression solution excluding these
objects is plotted as a dotted line and has a slope within 0.8 \% of
unity.  }
 \label{fig:flxflx}
\end{figure}

\subsection{Previous \hi detections}

Previous measurements of the \hi content of our detections were
available within NED and LEDA (and the references tabulated there) in
132 of 155 cases. We plot our flux densities against the LEDA values in
Fig.~\ref{fig:flxflx}. The \hi fluxes tabulated in LEDA correspond to
weighted averages of all previously published values. Those cases
marked in the last column of Table~\ref{tab:higal} as being possibly
confused in our telescope beam (having one or more cataloged companion
galaxies within 30~arcmin radius and 400~\kms) are plotted separately
as the open circles in Fig.~\ref{fig:flxflx}.  The distribution is
consistent with essentially the same absolute flux scale for the
isolated galaxies.  A linear regression solution (fit to the linear
fluxes rather than their logarithm) is overlaid on the data in
Fig.~\ref{fig:flxflx} and has a slope of 0.992, corresponding to a mean
flux-scale discrepancy of less than 0.8\%. The confused galaxies of our
sample show both a larger scatter and a systematic trend for an excess
\hi flux detected in our larger telescope beam.

\subsection{\hi in galaxy environments}

An important difference between the flux measurements reported here and
those in the literature is the large effective beam size of our
survey. Indeed, compared to the 3.1$\times3.7^\prime$ {\sc FWHM} beam
of the upgraded Arecibo telescope, our 46$\times49^\prime$ beam has a
200 times greater solid angle.  The linear {\sc FWHM} diameter of our
survey beam varies from about 70~kpc at the nearest galaxy distance of
5~Mpc to more than 1~Mpc at the furthest galaxy distance of 75~Mpc.

\begin{figure*}
\centering
 \begin{tabular}{cc}
 \resizebox{85mm}{!}{\includegraphics{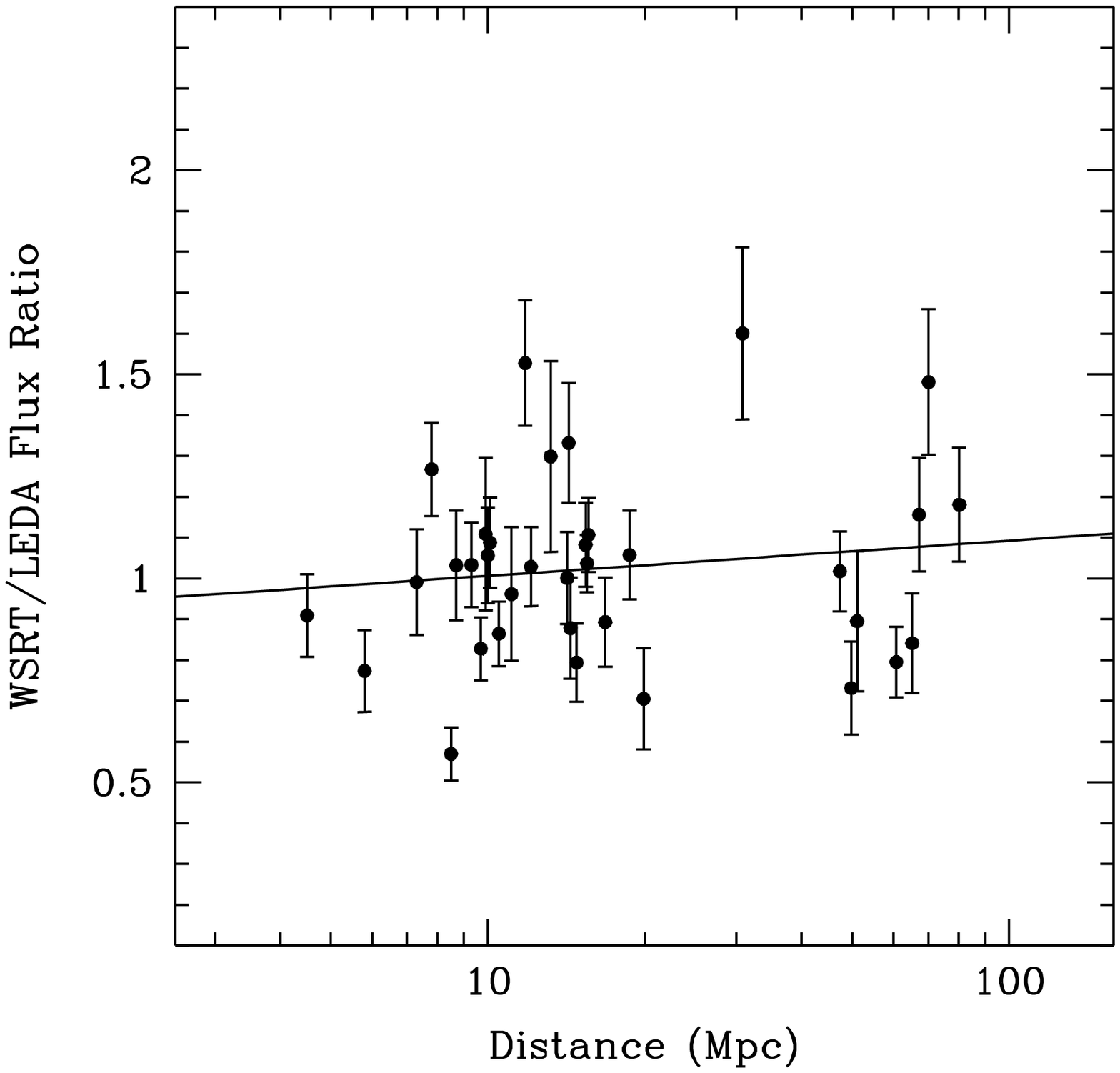}} &
 \resizebox{85mm}{!}{\includegraphics{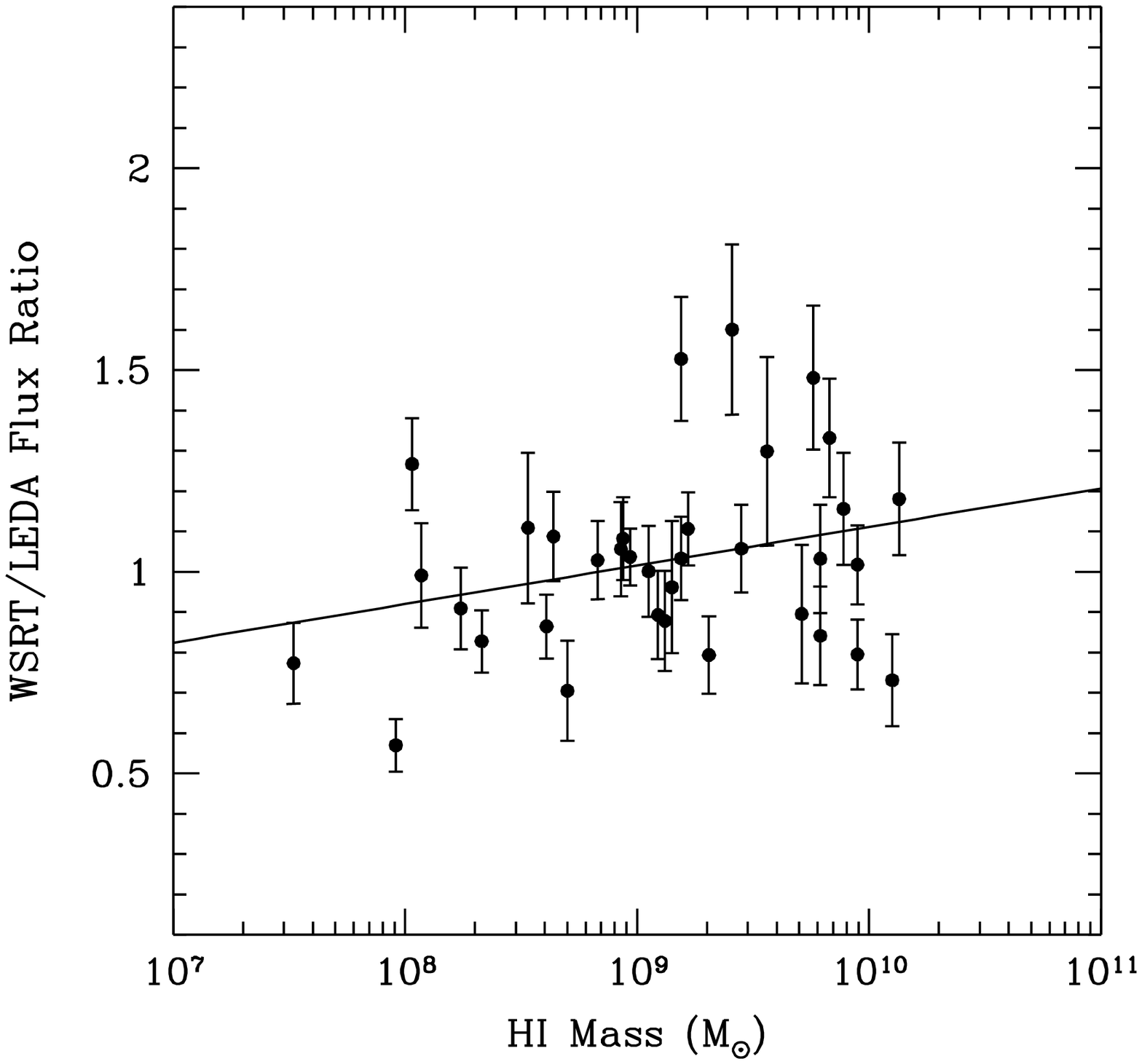}} \\
 \end{tabular}
 \caption{ Plots of our survey integrated \hi flux relative to those
measured previously as function of distance (left) and
\hi mass (right) for apparently isolated galaxies. Only those galaxies
with no cataloged companions within a radius of 30~arcmin and 400~\kms\ 
are plotted. Linear regression solutions are overlaid. }
 \label{fig:ratdm}
\end{figure*}

Given our larger beam area it is interesting to search for any
systematic increase in the \hi we detect relative to what has been
detected previously in a smaller beam. In the first instance we plot
the ratio of our survey \hi flux relative to that tabulated by LEDA as
function of distance and \hi mass in Fig.~\ref{fig:ratdm}. Only those
galaxies for which the flux ratio had a signal-to-noise greater than 5
are plotted, after taking account of the uncertainties in both our
value and that tabulated by LEDA.

The data-points are relatively few in number and quite noisy.
Essentially no excess flux is seen as function of distance (the
distribution has a correlation coefficient, $r$~=~0.080 and Student's
$t$~=~0.269, corresponding to a probability of significance, $P$~=~60\%),
while a weak trend of excess detected \hi flux may be present as
function of \hi mass ($r$~=~0.195, $t$~=~0.946, $P$~=~85\%). The linear
regression solutions with equal weights given to all points are
overlaid in both cases.

\begin{figure*}
\centering
 \begin{tabular}{cc}
 \resizebox{85mm}{!}{\includegraphics{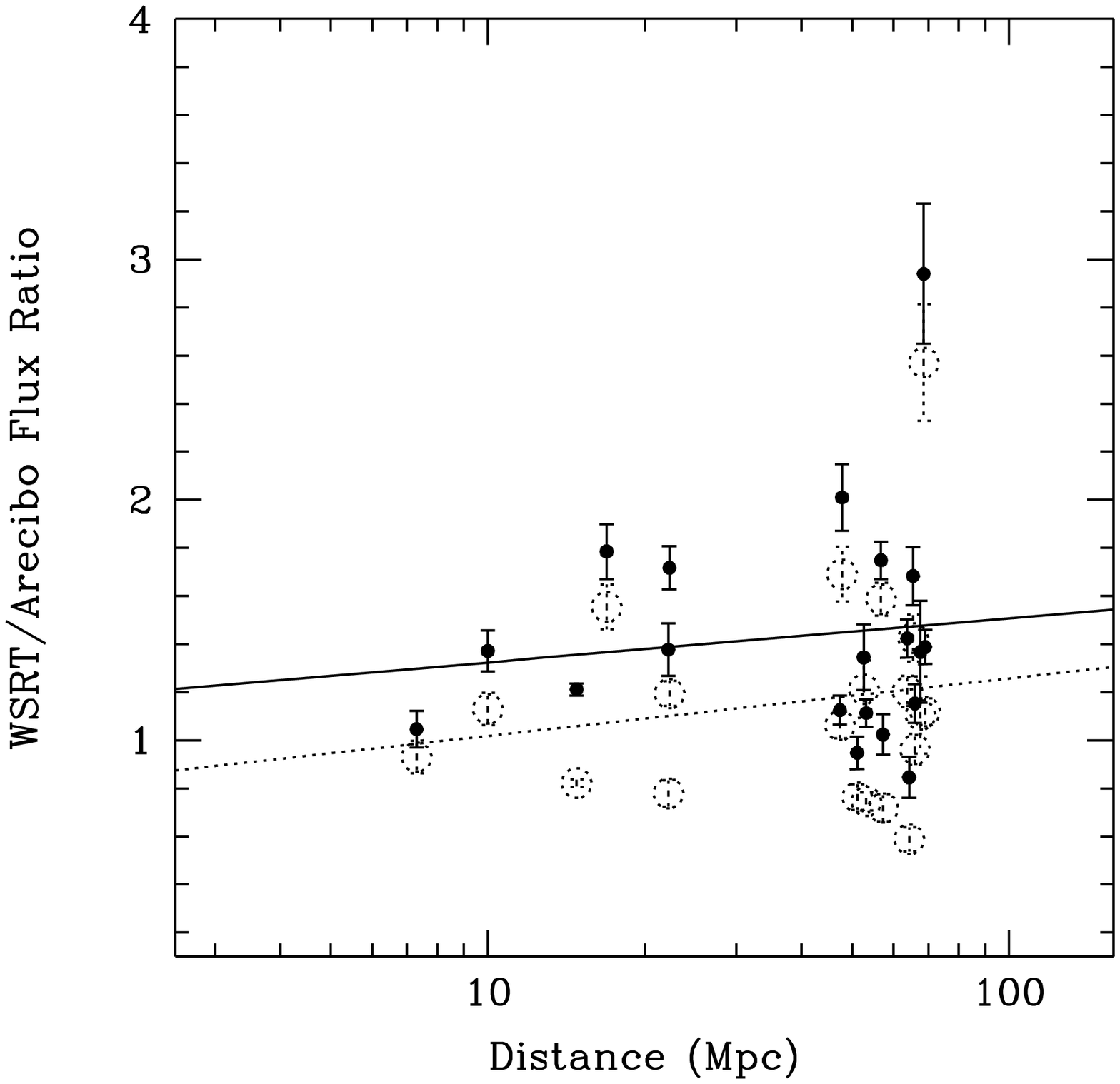}} &
 \resizebox{85mm}{!}{\includegraphics{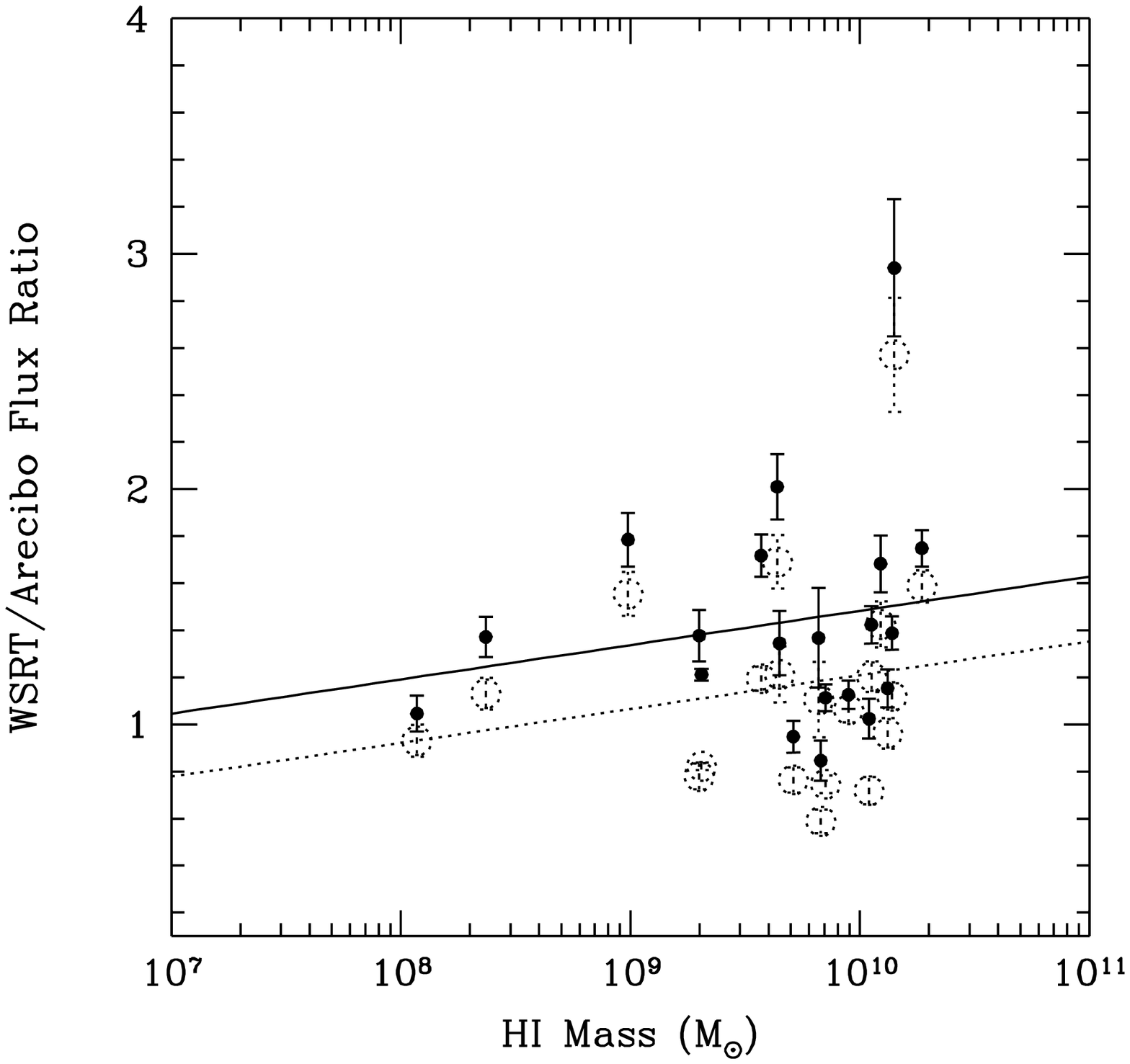}} \\
 \end{tabular}
 \caption{ Plots of our survey integrated \hi flux relative to that
measured with the Arecibo telescope as function of distance (left) and
\hi mass (right) for apparently isolated galaxies. Our survey beam
probes a 200 times larger solid angle that varies in linear diameter
from about 70~kpc at 5~Mpc to more than 1~Mpc at 75~Mpc. Only those
galaxies with no cataloged companions within a radius of 30~arcmin and
400~\kms\ are plotted. The {\it observed\ } flux ratios are plotted as
filled circles with error bars, while dotted open circles are used for
flux ratios that have been corrected in an approximate way for
telescope pointing errors and estimated source extent as seen with
Arecibo. Linear regression solutions are overlaid.}
 \label{fig:ratdma}
\end{figure*}

The LEDA data has been obtained from a wide variety of sources with a
corresponding variety in both beam size and calibration strategy. To
eliminate these variables from the flux comparison we also plot the
ratio of our survey \hi flux relative to that measured previously with
the Arecibo telescope as function of distance and \hi mass in
Fig.~\ref{fig:ratdma}.  The Arecibo data is taken from the
Pisces-Perseus supercluster survey (Giovanelli \& Haynes \cite{giov85},
Giovanelli et al. \cite{giov86}, Giovanelli \& Haynes \cite{giov89},
Wegner et al. \cite{wegn93} and Giovanelli \& Haynes \cite{giov93}). We
have only plotted the data for apparently isolated galaxies; those with
no cataloged companions lying within a radius of 30~arcmin and
400~\kms\ as indicated in the last column of Table~\ref{tab:higal}. The
flux ratio calculated from the {\it observed\ } Arecibo \hi flux is
plotted as the filled circles with error bars. A corrected Arecibo \hi
flux is also listed in these references, in which approximate
corrections are applied for telescope pointing errors, a model of the
finite angular extent of the target galaxies and the likely effect of
\hi self-absorption. Since the correction for \hi self-absorption
should affect both our measurements to an equal degree, we undo this
correction to the Arecibo fluxes before calculating the ratio. We plot
the {\it corrected\ } Arecibo to WSRT flux ratio as the dotted open
circles in the figure. The error bars in the corrected flux ratio do
not take account of the uncertainties in the correction procedure.

While the data-points are even fewer in number, they may suggest a
weak systematic excess of detected \hi flux within our larger survey
beam.  For the flux ratio as function of distance the distribution is
characterized by $r$~=~0.120, $t$~=~0.404 and $P$~=~65\%, while as
function of mass $r$~=~0.183, $t$~=~0.619 and $P$~=~73\%. The corrected
flux ratio still shows an excess for the majority of sources (with
$r$~=~0.162, $t$~=~0.475 and $P$~=~68\% as function of distance and
$r$~=~0.187, $t$~=~0.819 and $P$~=~78\% as function of mass), although
the incidence of several corrected flux ratios significantly less than
unity casts some doubt on the general reliability of the correction
procedure.

\subsection{New \hi detections}

Twenty-three of the objects listed in Table~\ref{tab:higal} have been detected
in \hi for the first time in our survey. We comment
briefly on each of these objects below.

J2202+4838, at ($l,b$)~=~(96.5,$-$5.4) is the only object in our
8$\sigma$ sample with no apparent optical counterpart within a
30~arcmin search radius in the second generation DSS images. The
predicted extinction in this direction is moderate but not extreme,
A$_B$~=~1.32~mag (Schlegel et al. \cite{schl98}). 

UGC~11923, classified merely as type ``S'', has received relatively
little study, no doubt due in part to it's position
($l,b$)~=~(94.9,$-$9.3) and consequently relatively high extinction,
A$_B$~=~1.74~mag (Schlegel et al. \cite{schl98}). This is a moderately
gas-rich system, with log(M$_{HI}$)~=~10.06.

UGC~11929 is a little-studied S0 galaxy with IRAS fluxes in the 60 and
100$\mu$m bands of 1.3 and 2.7 Jy.

KKR~71 is a nearby irregular system that has received little study.

CGCG~514-098 is a moderately distant unclassified galaxy with IRAS
fluxes in the 60 and 100$\mu$m bands of 1.3 and 3.3 Jy.

CGCG~476-100 is an unclassified galaxy with no previous red-shift
determination.

UGC~64 is a nearby low mass system with a peculiar optical morphology,
possibly suggestive of tidal interaction (Vorontsov-Velyaminov
\cite{voro77}).

And~IV was only recently recognized (Ferguson et al. \cite{ferg00} as a
low mass dwarf in the background (at 5 to 8~Mpc distance) of M~31,
rather than being closely associated with M~31 as had been thought
previously. In fact, together with UGC~12894, UGC~64 and UGC~288,
And~IV forms a nearby filament of low-mass galaxies.

CGCG~501-016 is an unclassified galaxy that was host to SN 1995am with
IRAS fluxes in the 60 and 100$\mu$m bands of 0.2 and $<$0.84 Jy.

CGCG~502-039 is an unclassified galaxy with IRAS fluxes in the 60 and
100$\mu$m bands of 1.1 and 1.74 Jy.

CGCG~521-039 is an unclassified galaxy which has received little study.

J0136+4759 appears to be associated with a previously uncataloged LSB
galaxy at ($\alpha_{2000},\delta_{2000}$)~=~(01:36:40,+48:03:40) as
illustrated in Fig.~\ref{fig:lgco}. The optical galaxy was presumbaly
not recognized previously due to it's close proximity to a moderately
bright foreground star.

NGC~661 is classified as an E+, with the UGC noting a diffuse companion
at 3.8 arcmin offset in PA 261 east of north. Chamaraux et
al. (\cite{cham87}) report a non-detection ($<$0.57~Jy-\kms) at the
optical position and red-shift measured in one Arecibo beam (3.9 arcmin
{\sc FWHM}), suggesting that our detection, with it's 9.4 arcmin
offset, may be due to either a companion of NGC~661 or tidal debris.

[ZBS97]~A31 was first detected in the AHISS survey (Zwaan et
al. \cite{zwaa97}), where it is noted as having log(h$^{-2}$M$_{\hi}$)=8.97
Our detection centroid is offset by about 10 arcmin and appears to be
significantly more massive. 

NGC~780 is an unclassified galaxy, with apparent stellar plumes
extending in several directions from the main galaxy body, suggesting a
recent merger remnant.

CGCG~538-034 is classified as S0 and has IRAS fluxes in the 60 and
100$\mu$m bands of 3.0 and 3.0 Jy

UGC~01830 is classified as an SB0/a and has IRAS fluxes in the 60 and
100$\mu$m bands of 1.2 and 3.1 Jy. The UGC notes: ``Very
compact core, double ring halo.''

IC~1815 is classified as an SB0, and has a diffuse stellar halo.

J0239+3015 appears to be associated with the NED object [VR94]
0236.9+3003 with tabulated photometry by Vennik \& Richter
\cite{venn94}, but without a previous red-shift determination.

J0240+4221=IRAS 02371+4223 is an unclassified galaxy with IRAS fluxes
in the 60 and 100$\mu$m bands of 0.7 and 1.3 Jy.

UGC~2172 represents the lowest integrated \hi flux detection of our
sample with only 3.1$\pm$0.35~Jy-\kms, corresponding to
log(M$_{HI}$)~=~7.85. A previous unsuccessful search for \hi in this
galaxy by Schneider et al. (\cite{schn92}) was directed at the
incorrect velocity interval, since an optical red-shift only became
available in 1999.

NGC~1161 is classified as an S0 galaxy and has previously been searched
for \hi emission by Haynes et al. (\cite{hayn90} using the Green Bank
300 ft telescope down to an {\sc rms} sensitivity of 3.5 mJy/Beam over
5.5~\kms\  velocity channels. The extreme positional offset of our
detection (13.5 arcmin corresponding to 100~kpc) coupled with the
GB300' non-detection suggest that we are likely detecting either
tidally stripped gas at large radii or a gas-rich companion rather than
NGC~1161 itself.

HFLLZOA G144.00-08.53 is an obscured system (A$_B$~=~2.05~mag (Schlegel
et al. \cite{schl98}), classified as a dE, although the compact central
concentration is surrounded with more diffuse stellar emssion in the
DSS. No previous red-shift is available for this source.

\subsection{The \hi mass function}

An important application of blind \hi surveys is the characterization
of the general population of neutral gas-rich objects without the
inevitable bias associated with the study of an optically-selected
sample. The 155 \hi detections which follow from our 8~$\sigma$ limit on
integrated \hi flux, form a relatively small, but moderately complete
sample with which to characterize the population. Indeed, extensive
simulations with synthetic sources in a comparable survey led Rosenberg
\& Schneider (\cite{rose02}) to conclude that above an effective signal
to noise ratio of 8, their sample of \hi selected objects was
essentially complete.  Since our sample is limited in flux density,
which scales as the inverse square of distance for a given \hi mass, it
is clear that we sample a very different survey volume at low \hi mass
relative to high.  For example, an object with log(M$_{\rm HI}$)~=~7.5
and W$_{20}$~=~40~\kms\  can only be detected out to D~=~5.4~Mpc, while
Galactic \hi emission extends out to about V$_{Hel}$~=~200~\kms,
corresponding to V$_{LGSR}~\sim$~350~\kms\  and D~=~4.7~Mpc, leaving a
possible detection volume of only some 10~Mpc$^3$. On the other hand an
object with log(M$_{\rm HI}$)~=~10. and W$_{20}$~=~300~\kms\  can be
detected beyond the edge of our survey volume at about D~=~88~Mpc,
corresponding to more than 10$^5$~Mpc$^3$. When comparing our
detections in different \hi mass bins it becomes important to consider
whether the average space density of galaxies is actually uniform
within our survey volume.

\begin{figure*}
\centering
 \includegraphics[width=17cm]{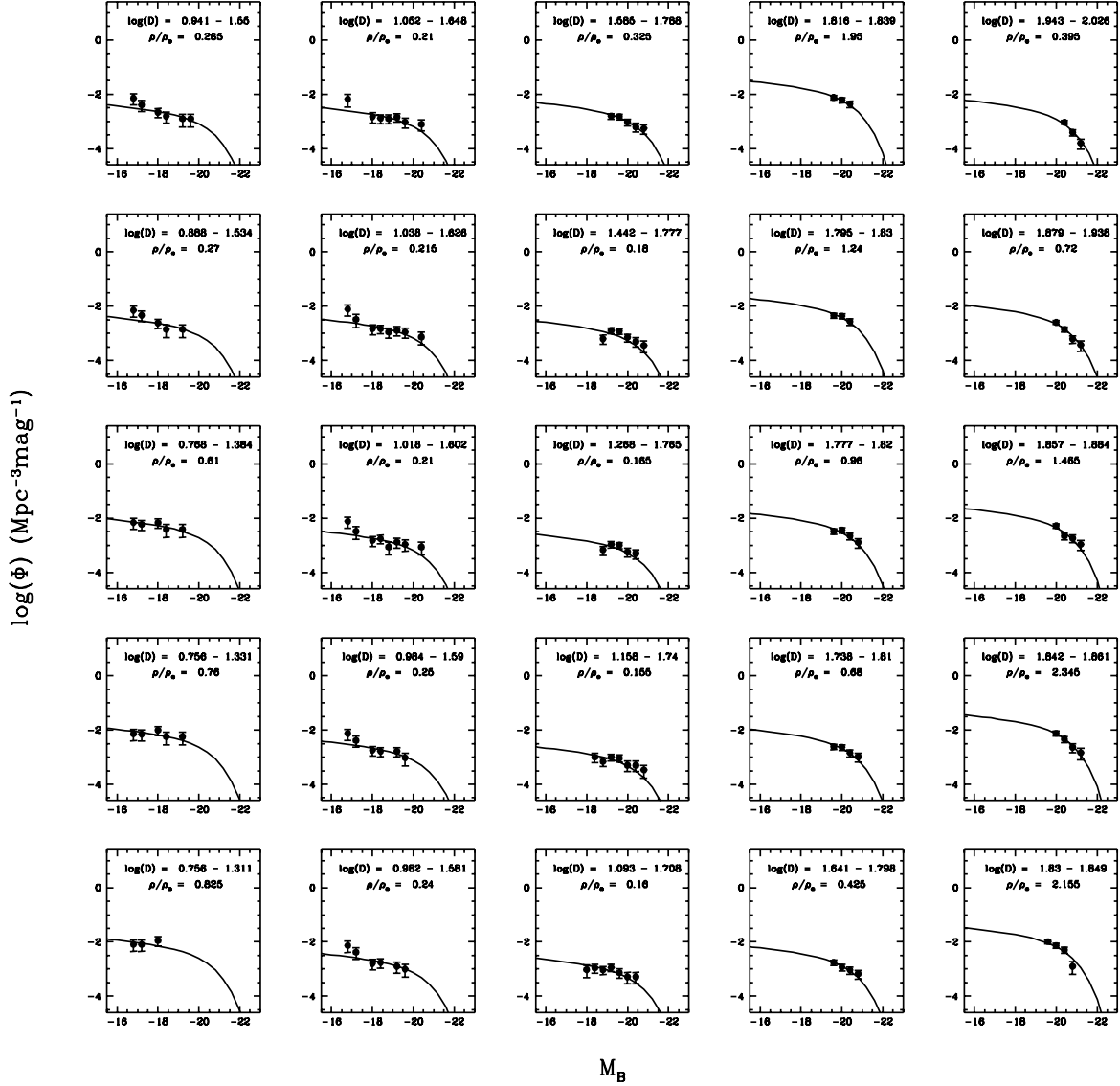}
 \caption{ Optical luminosity distributions over the survey region in the
indicated intervals of log(D). The solid curves are Schechter
functions with M$_B^*-5log_{10}h ~=~-19.66$,
$\alpha~=~-1.21$ and the indicated density relative to
$\Phi_*~=~1.61\times10^{-2}h^3$. The ``standard'' optical luminosity
function parameters are taken from Norberg et al. (\cite{norb02}).}
 \label{fig:lfdist}
\end{figure*}

The wedge diagrams of our \hi detections shown in Fig.~\ref{fig:wedge}
already gave some indication for non-uniformity of the space density
along the line-of-sight, despite the rather substantial solid-angle of
our survey. A moderate galaxy density is seen between 5 and 15~Mpc,
followed by an apparent void and subsequently another enhancement.
Since this is difficult to quantify directly on the basis of our
limited number of detections, we have instead considered the space
density of optically cataloged galaxies as a function of recession
velocity. We extracted from the LEDA database all galaxies of known
LGSR recession velocity (greater than $+250$ \kms) and integrated
B-band magnitude within the spatial and velocity boundaries of our
survey region. A total of 1774 galaxies are cataloged within LEDA
within our survey boundaries. We adopt an approximate completeness
limit of m$_{BT}$~=~14.9, based on the turn-over in the cumulative
distribution function. 
Constraining the selection to m$_{BT} <$~14.9, we retain about
$N_{Gal}$~=~530 optically cataloged galaxies in our survey
volume. After sorting these galaxies by distance, they have been
divided into 25 overlapping sub-samples with sample populations varying
linearly from a minumum of $N_{Pop}$~=~40 galaxies at the nearest
distances to $N_{Pop}^\prime$~=~100 galaxies at the maximum distance.
Adjoining sub-sample populations share more than half of their
membership to insure sufficient sampling of the density variation with
distance. (This has been accomplished by choosing the start index,
$i~=~1\dots N_{Gal}$, of sub-sample, $j~=~1\dots25$, in the ordered
galaxy list using the prescription $i =
(N_{Gal}-N_{Pop}^\prime)^{((j-1)/24)^{0.8}}$.)  We plot the optical
B-band luminosity distributions of these 25 sub-samples in
Fig.~\ref{fig:lfdist}. The relevant intervals of log(D) are indicated
at the top of each panel.  Also plotted is a ``standard'' luminosity
function taken from Norberg et al. (\cite{norb02}),
$$ d\Phi({\rm M})/d{\rm M} = \Phi_* {\rm ln}(2.512) ({\rm L}/ {\rm
L_*})^{\alpha+1} {\rm exp}(-{\rm L/L_*}),$$ with L$_*$ corresponding to
M$_B^*-5log_{10}h ~=~-19.67$, $\alpha~=~-1.21$ and the indicated
density relative to $\Phi_*~=~1.61\times10^{-2}h^3$, which they derive
from more than 10$^5$ galaxies in the 2dF red-shift survey. As can be
seen in the panels of the figure, the luminosity distributions are
moderately complete, with only the occasional down-turn in the lowest
luminosity bin. All of these distributions can be reasonably
well-described by the same 'standard' luminosity function, where the
only permitted variable in a $\chi^2$ minimization was the galaxy space
density relative to the global average value of
$\Phi_*~=~1.61\times10^{-2}h^3$ found by Norberg et al. Values of
$\Delta\chi^2$~=~1, corresponding to 1$\sigma$ errors in the
over-density, varied from about 20\% at the smallest distances to 10\%
at the largest distances. These appear to be realistic error estimates
under the assumptions that the density is a smoothly varying
distribution in our survey volume and that the shape of the luminosity
function is not also a function of distance. The large degree of
overlap of the sub-samples in adjoining distance intervals does not
have an adverse impact on the error estimate, but simply insures
sufficient sampling of changes in the density with distance. 

\begin{figure}
 \resizebox{\hsize}{!}{\includegraphics{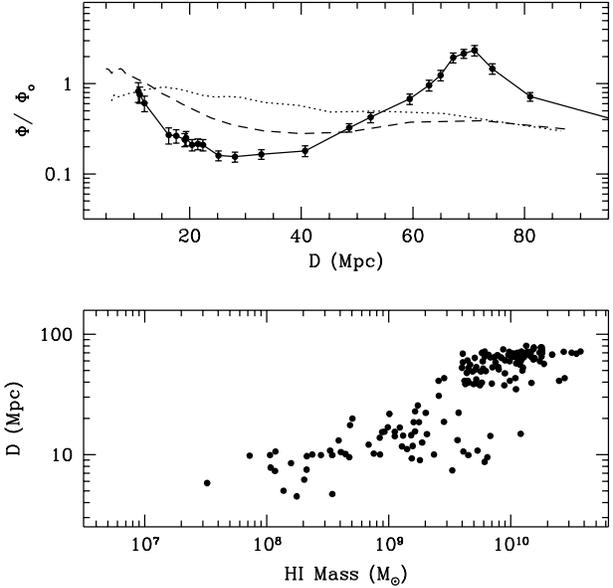}}
 \caption{The derived galaxy over-density in our survey volume as
function of distance (top) and the distribution of detected \hi masses
with distance (bottom). The over-density is derived from fits to the
the luminosity distribution of optical galaxy sub-samples as shown in
Fig.~\ref{fig:lfdist}. The dotted line in the top panel indicates the
same measure of optical galaxy over-density derived for the HIPASS
survey volume ($\delta < 0$), while the dashed line is that for the
northern hemisphere ($\delta > 0$).}
 \label{fig:lgsover}
\end{figure}

Our derived variation of galaxy density within our survey boundaries as
function of distance is plotted in the upper panel of
Fig.~\ref{fig:lgsover}, with the 1$\sigma$ error bars noted above. The local
galaxy density (within about 10~Mpc) is slightly below the standard 2dF
value. This plummets to some 20\% of the average in the void near
D~=~25~Mpc, slowly climbs to an overdensity centered at 70~Mpc and
subsequently declines. The impact of such a variation of density can be
judged in the lower panel of Fig.~\ref{fig:lgsover}, where our
detections are plotted as a function of \hi mass and distance.

\begin{figure}
 \resizebox{\hsize}{!}{\includegraphics{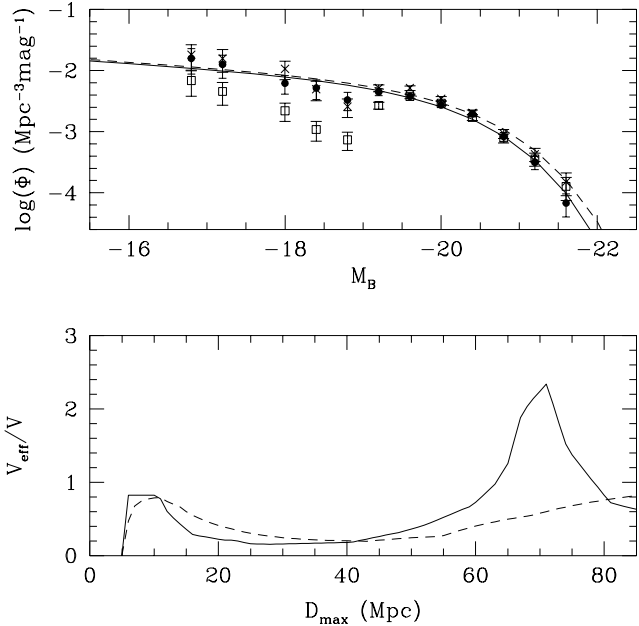}}
 \caption{ Distribution of effective survey volume (bottom) and
density-corrected luminosity function (top). The solid and dashed
curves in the lower panel contrast the ``discrete'' and ``integral''
formulations of the effective survey volume as function of the limiting
distance, $D_{max}$, under the simplifying assumption that most
detections occur near the limiting distance. The symbols in the top
panel give the observed (open square) and density-corrected luminosity
functions (filled circles for the discrete-, and crosses for the integral
formulations of V$_{eff}$) in our entire survey volume. The solid and dashed
curves are Schechter function fits to the density-corrected data from the
discrete and integral formulations. }
 \label{fig:veff}
\end{figure}

The suggested method of introducing density corrections in optical
luminosity functions or an HIMF (Saunders et al. \cite{saun90},
Rosenberg \& Schneider \cite{rose02}) is the calculation of a so-called
effective volume, V$_{eff}\propto\int_{D_{min}}^{D_{max}}\rho(D)D^2dD$,
to replace the physical volume
V$_{tot}\propto\int_{D_{min}}^{D_{max}}D^2dD$, where the integrals
extend over the entire distance range over which each source could have
been detected. We considered that a more straightforward method of
achieving the desired result, of a ``uniform density'' survey volume,
might be simply
V$_{eff}^\prime\propto\rho(D)\int_{D_{min}}^{D_{max}}D^2dD$, where the
density has been taken out of the integral and is only evaluated at the
distance of the detection in question. Although somewhat challenging to
compare these two formulations of V$_{eff}$ directly, this can be done
approximately by considering that the largest number of detections of
any distribution which is rising at it's faint end will be near the
limiting distance, $D_{max}$. With this simplifying assumption we
compare these two formulations in the lower panel of
Fig.~\ref{fig:veff}, where the ratio of effective to physical volume is
plotted as a function of $D_{max}$.  The ``discrete'' formulation of
V$_{eff}^\prime$ is plotted as the solid line and exactly traces the
distribution of over-density plotted in Fig.~\ref{fig:lgsover} after
normalization with the physical volume. The ``integral'' formulation of
V$_{eff}$ is plotted as the dashed line in Fig.~\ref{fig:veff}. This
formulation shows large systematic departures from the discrete one,
with peaks and troughs shifted to higher distances and having decreased
amplitude.  The density-corrected optical luminosity functions which
follow from the discrete and integral formulations are compared in the
upper panel of Fig.~\ref{fig:veff}. The open squares in this figure
indicate the accumulated (1/V$_{\rm tot}$) luminosity function over our
entire survey volume with no density correction, while the filled
circles and crosses are the density-corrected luminosity
functions. Both forms of density correction give a substantial
improvement in recovering the template luminosity function of Norberg
et al. (\cite{norb02}). The best $\chi^2$ fits (constrained only to
have the Norberg et al. power-law of $-$1.21) are overlaid in the
figure as the solid and dashed curves. The other Norberg et
al. Schechter function parameters are recovered to within the 1$\sigma$
errors in both cases, although the fit residuals are significantly
higher in the case of the integral formulation, leading to a higher
value of the reduced $\chi^2$~=~33.3, compared to $\chi^2$~=~12.5 for
the discrete formulation. In view of the much lower reduced $\chi^2$ of
the discrete density-correction method we have chosen to utilize this
approach in correcting our HIMF.

\begin{figure*}
\centering
 \includegraphics[width=17cm]{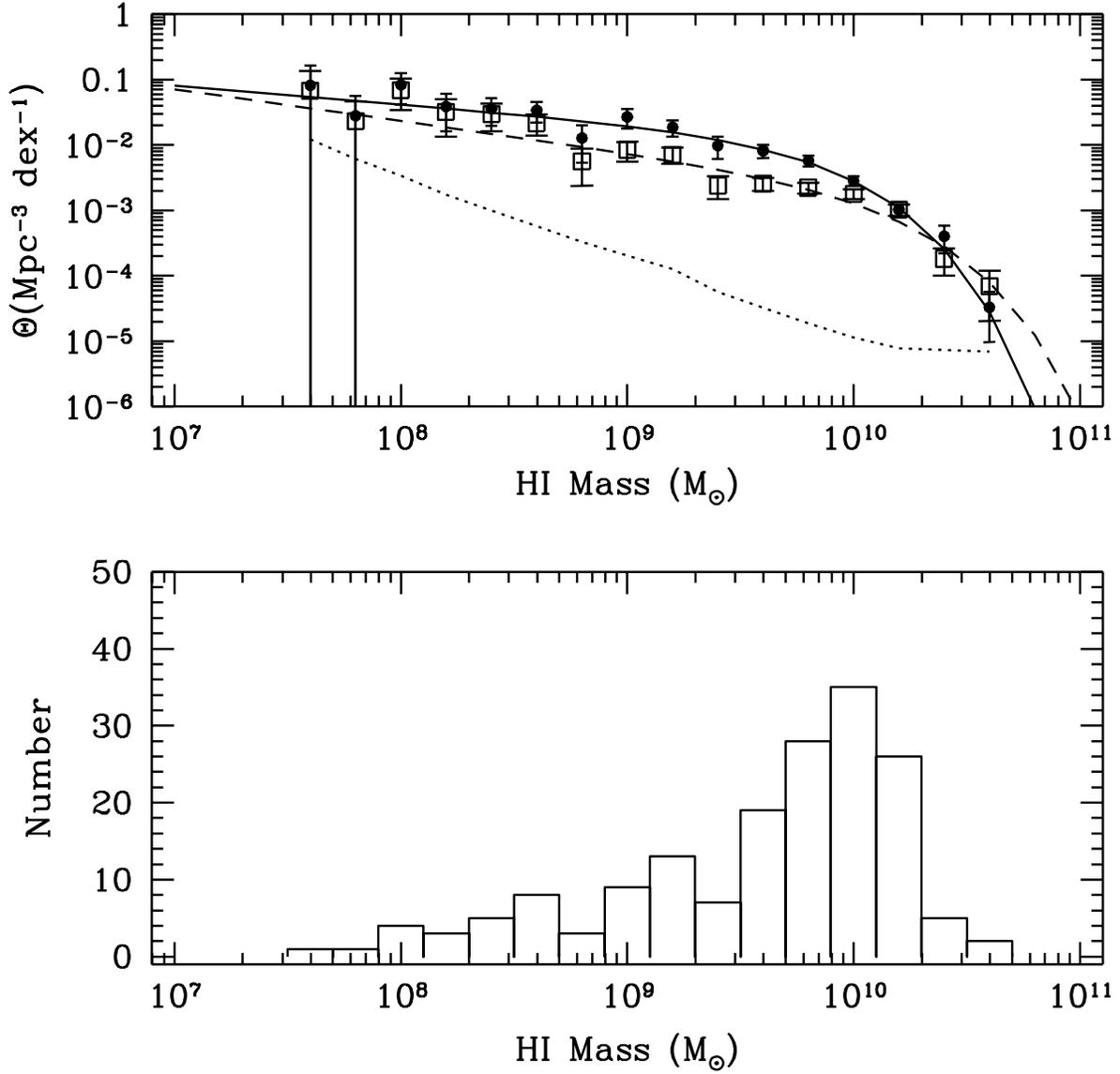}
 \caption{Distribution of detected \hi masses (bottom) and the derived
HIMF (top). The open squares are derived from a straightforward
application of 1/V$_{tot}$, while the filled circles have been
corrected for the variation of galaxy density with distance. The dashed
and solid curves are the best-fitting Schechter functions with
log(M$_*)~=~10.2$, $\Theta_*~=~9.5\times10^{-4}$ and $\alpha~=~-1.48$
for 1/V$_{tot}$, and log(M$_*)~=~9.85$, $\Theta_*~=~55.\times10^{-4}$
and $\alpha~=~-1.28$ after the density correction. The dotted curve
represents the reciprocal of the limiting survey volume as function of
mass.  }
 \label{fig:lgshimf}
\end{figure*}

We are now in a position to determine the HIMF from our galaxy
detections. The histogram of galaxy detections as function of \hi mass
is shown in the bottom panel of Fig.~\ref{fig:lgshimf}, while the
corresponding mass functions are shown in the top panel of the
figure. The mass function was calculated by accumulating each galaxy
detection divided by the maximum volume out to which that object would
still have satisfied our detection criterion of an 8$\sigma$ integrated
flux density, normalized as usual to a binwidth of one dex in \hi
mass. We have explicitly taken account of the variation of survey
sensitivity with recession velocity (as illustrated in
Fig.~\ref{fig:specrms}) in the calculation of the limiting survey
volume for each detected galaxy.  This corresponds to the classical
(1/V$_{\rm tot}$) method developed by Schmidt (\cite{schm68}). Open squares
are used in the figure to indicate the (1/V$_{\rm tot}$)
datapoints, while the filled circles have also been corrected for the
variation of density with distance in our survey region (as plotted in
Fig.~\ref{fig:lgsover}). Based on our comparison of the
``discrete'' and ``integral'' formulations of density-correction
discussed above, we chose to apply the method of discrete density
correction, in which
V$_{eff}^\prime\propto\rho(D)\int_{D_{min}}^{D_{max}}D^2dD$. We have
also considered density correction of the HIMF with the integral
formulation of V$_{eff}$ and find similar results, but with a reduced
$\chi^2$ value of the best-fitting Schechter function almost three
times as large ($\chi^2_{min}$~=~15 compared to 6.3). 

We have included all 155 of our blind \hi detections in the mass
function, although in those cases where our flux density measurement of
the primary optical ID had a large uncertainty, either due to possible
confusion by nearby companions (which applies to 70 of our detections
as noted in the last column of Table~\ref{tab:higal}) or to a recession
velocity near V$_{Hel}~\sim~2800$~\kms\  (the transition in our velocity
coverage from the lower to the upper band of 20~MHz width) we have used
the LEDA flux values to calculate the \hi mass rather than our own. In
addition, we have considered all cataloged galaxies that could
contribute to source confusion within our survey beam in the vicinity
of our 70 confused detections. Those confusing galaxies that had
tabulated LEDA fluxes sufficient to satisfy our 8~$\sigma$ limit on
integrated \hi flux, were also accumulated in the HIMF. This
consideration led to an additional 14 nearby companion galaxies being
incorporated into the mass function. The total number of galaxies
contributing to our HIMF is 169. 

The error bars of the datapoints in Fig.~\ref{fig:lgshimf} are
determined solely by the square-root of the \hi mass-bin occupancy,
which leads to errors that vary between about 20 and 100\%. These
errors dominate the error budget since they exceed the random erors
associated with the density-correction by a factor of between 2 and 5.

\begin{figure*}
\centering
 \begin{tabular}{ccc}
 \resizebox{55mm}{!}{\includegraphics{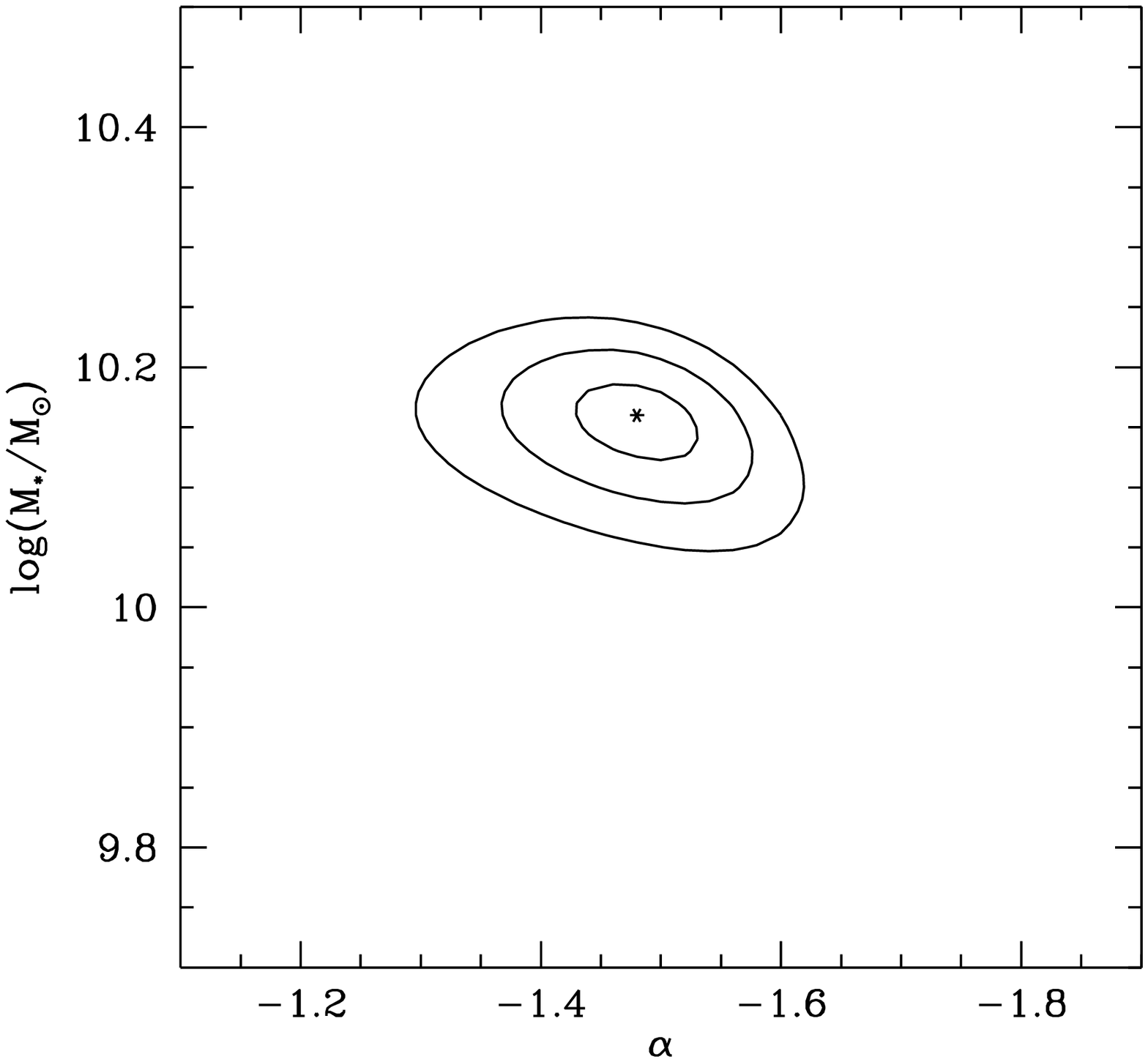}} &
 \resizebox{55mm}{!}{\includegraphics{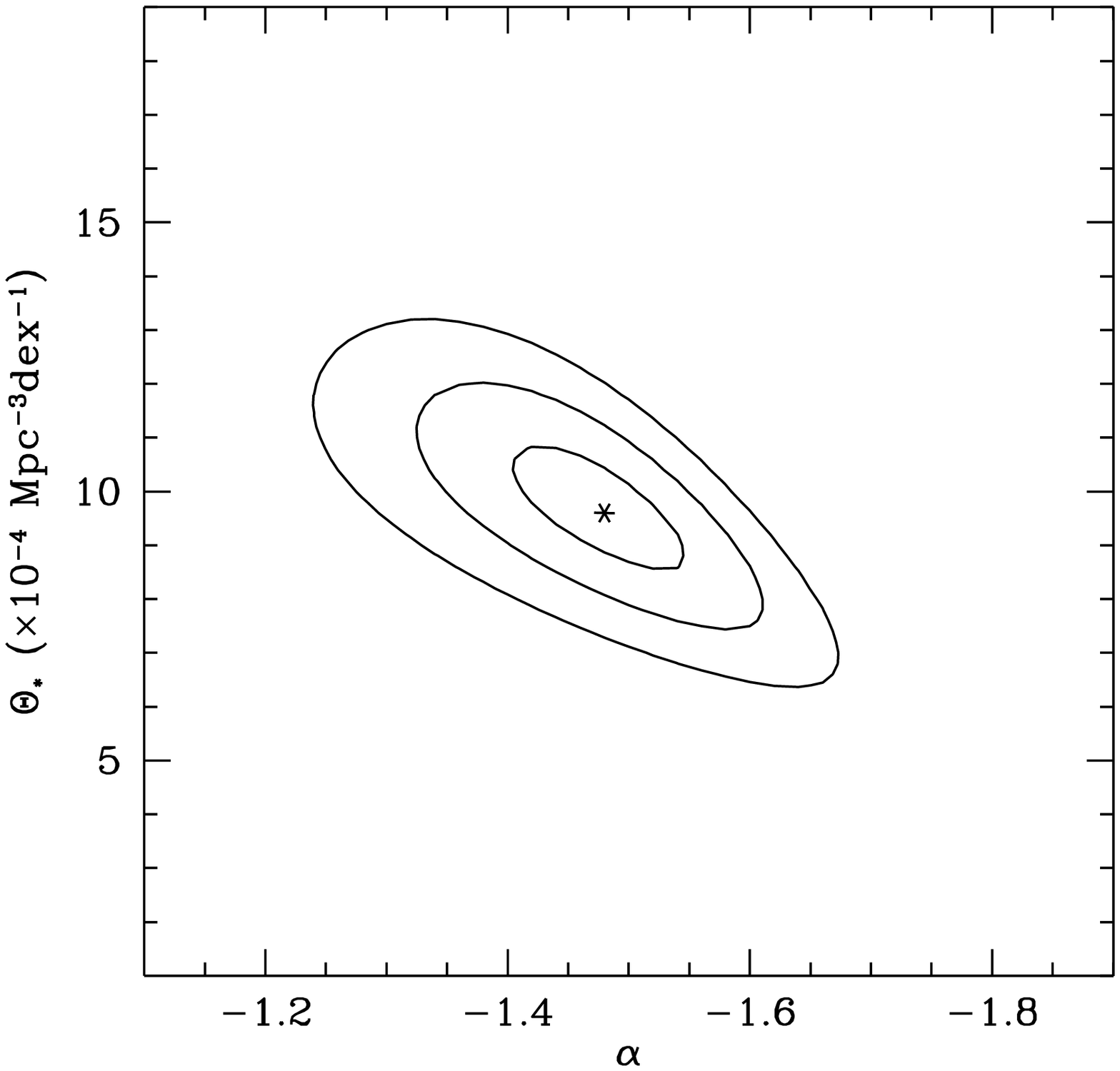}} &
 \resizebox{55mm}{!}{\includegraphics{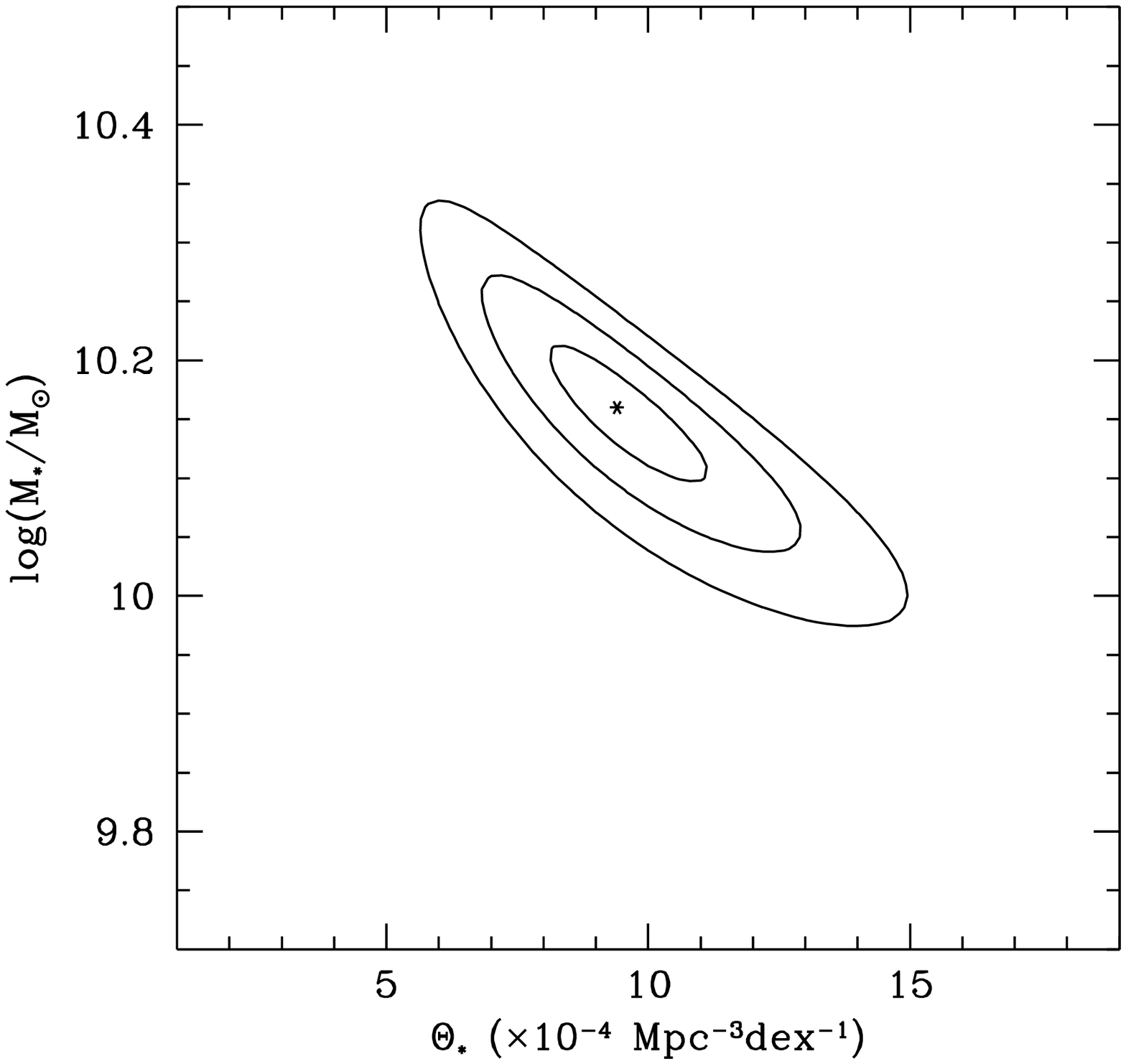}} \\
 \resizebox{55mm}{!}{\includegraphics{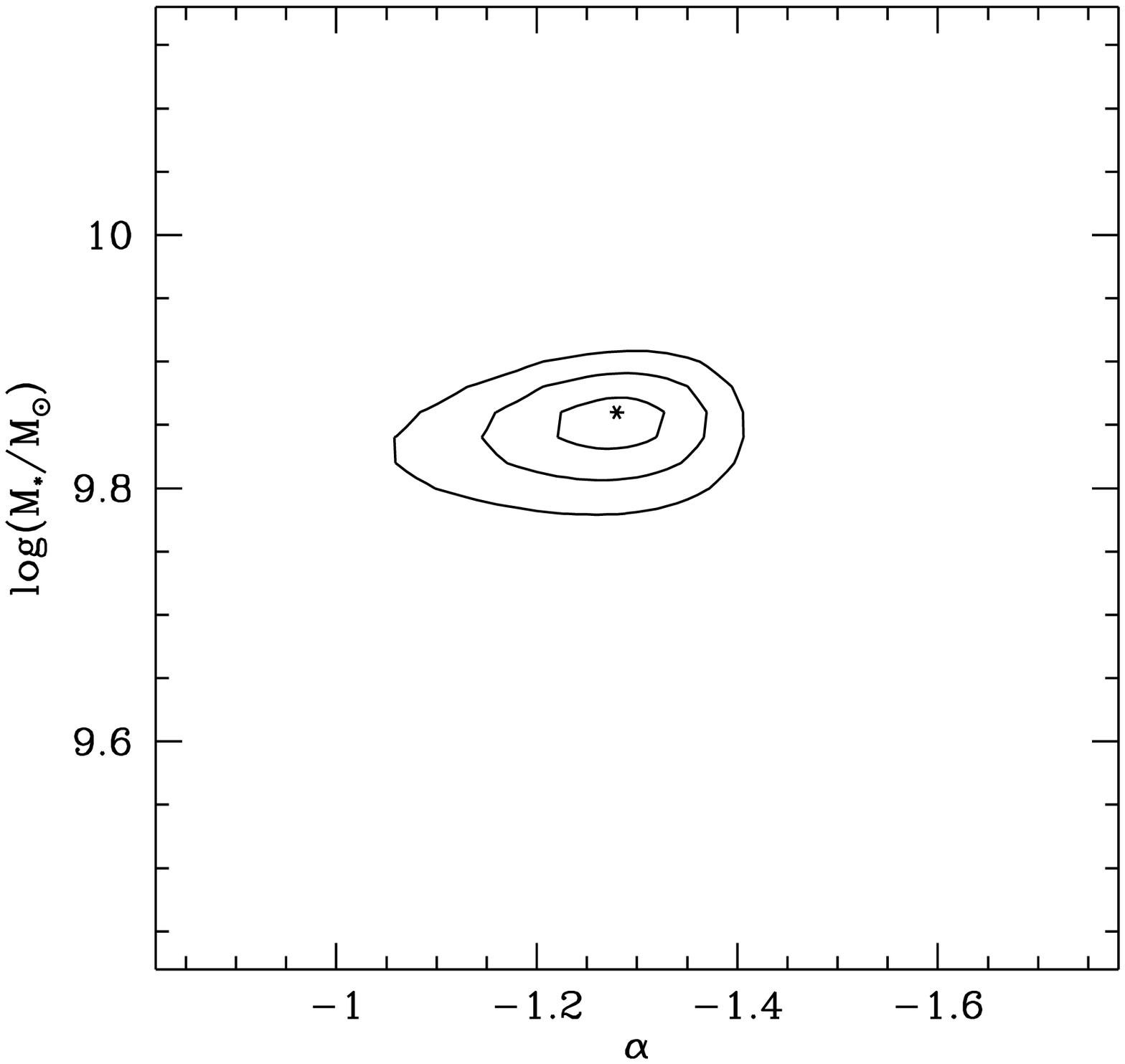}} &
 \resizebox{55mm}{!}{\includegraphics{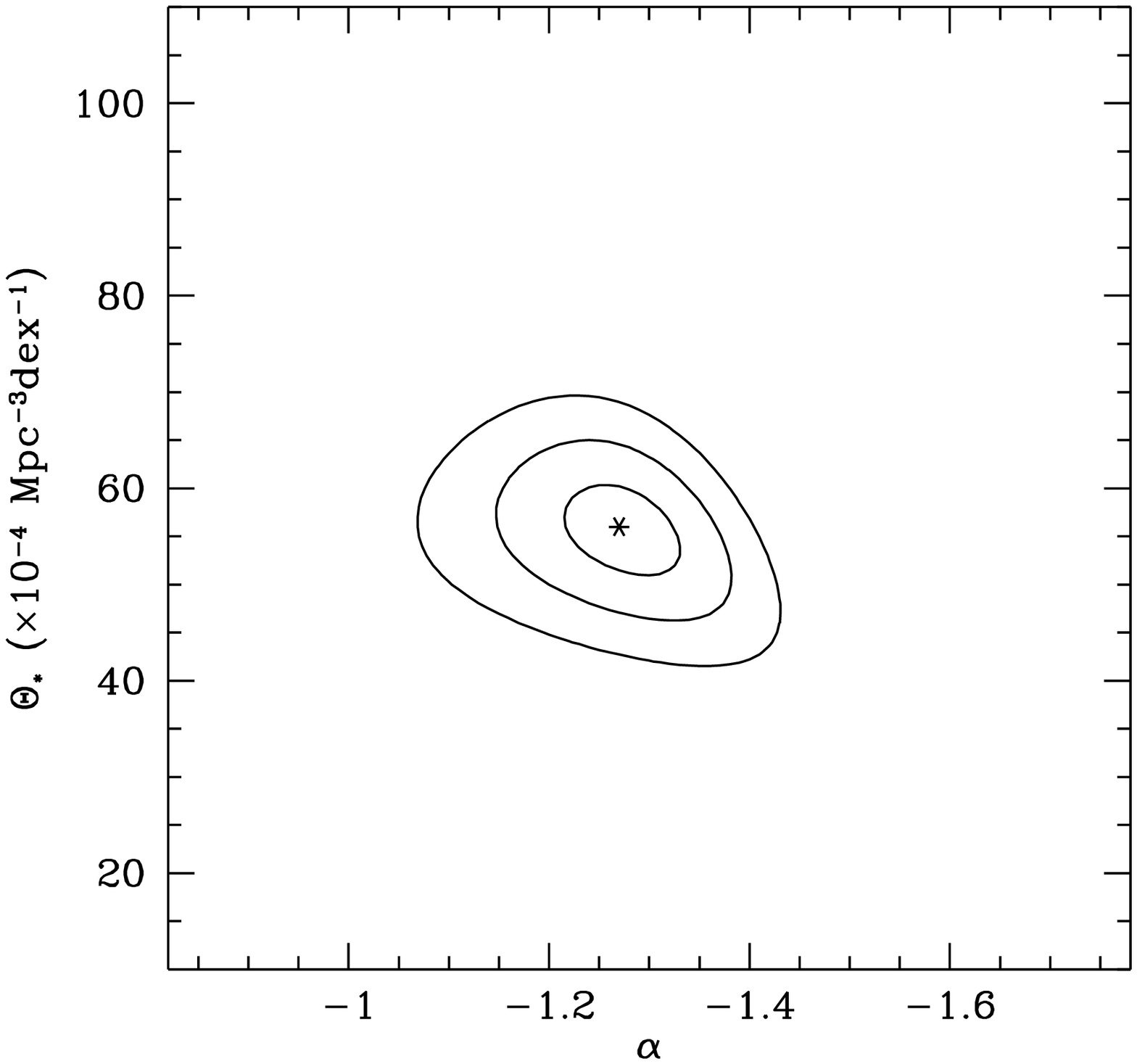}} &
 \resizebox{55mm}{!}{\includegraphics{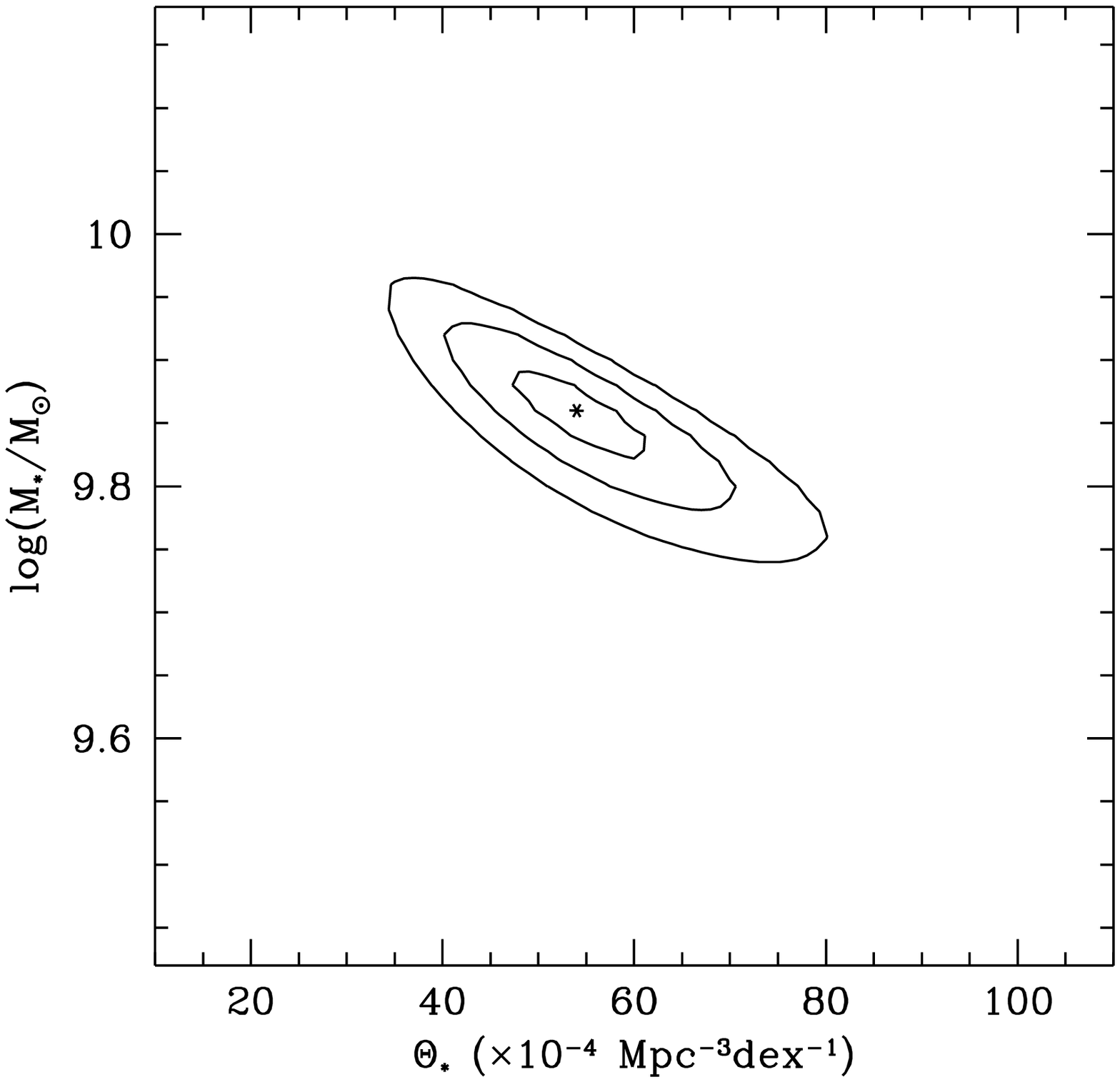}} \\
 \end{tabular}
 \caption{ Contours of $\chi^2$ for pairs of our fit parameters are
shown both before (top row) and after (bottom row) correction for
galaxy density variations with distance. Contours are drawn at
$\Delta\chi^2$~=~1, 4 and 9 corresponding to 1, 2 and 3$\sigma$ for one
degree of freedom. The third parameter is kept fixed at the
best-fitting value when plotting each pair.  }
 \label{fig:chipar}
\end{figure*}

The dotted line in Fig.~\ref{fig:lgshimf} represents the approximate
inverse search volume of our survey as function of mass, where we have
assumed a relationship between \hi mass and linewidth of the form:
W$_{20}~=~0.16$~M$_{\rm HI}^{1/3}$~\kms, for M$_{\rm HI}$ in solar units.
The best-fitting Schechter functions of the form:
$$ \Theta({\rm M}) = \Theta_* {\rm ln}(10) ({\rm M_{HI}} / {\rm
M_*})^{\alpha+1}{\rm exp}(-{\rm M_{HI}/M_*}),$$ are over-laid on the
data points. The straightforward (1/V$_{\rm tot}$) points are best-fit
with log(M$_*$)~=~10.15$\pm0.1$, $\Theta_*$~=~9.5$\pm3\times10^{-4}$
and $\alpha$~=~$-$1.5$\pm0.1$, indicated by the dashed line in
the figure. The best-fit values after correction for galaxy density as
given by the upper panel of Fig.~\ref{fig:lgsover} are
log(M$_*$)~=~9.85$\pm0.07$, $\Theta_*$~=~55$\pm15\times10^{-4}$ and
$\alpha$~=~$-$1.28$\pm0.1$, indicated by the solid line in the
figure. Contours of $\chi^2$ for pairs of our fit parameters are shown
in Fig.~\ref{fig:chipar} both before and after correction for galaxy
density variations with distance. The contours are drawn at
$\Delta\chi^2$~=~1, 4 and 9 corresponding to 1, 2 and 3$\sigma$ for one
degree of freedom. In each plot, the third parameter is kept fixed at
the best-fitting value.  From the $\chi^2$ contours it is clear that
the solutions for log(M$_*$) and $\alpha$ are well seperated, while
combinations involving the galaxy density become somewhat degenerate.

For comparison the HIMF derived by Zwaan et al. (\cite{zwaa03}) from
the HIPASS Bright Galaxy Catalog, based on 1000 \hi selected galaxies
in the $2\pi$ steradians below $\delta=0$, has
log(M$_*$)~=~9.79$\pm0.06$, $\Theta_*$~=~86$\pm21\times10^{-4}$ and
$\alpha$~=~$-$1.30$\pm0.08$. Although completely at odds with our
(1/V$_{\rm tot}$) values, good agreement is apparent between
these values and our own after application of the galaxy density
correction. Similar considerations apply to HIMF derived by Rosenberg
\& Schneider (\cite{rose02}) from the Arecibo Dual-Beam Survey (ADBS)
who find log(M$_*$)~=~9.88, $\Theta_*$~=~58$\times10^{-4}$ and
$\alpha$~=~$-$1.53, although with somewhat poorer agreement in the
faint end slope.

Although our choice of a minimum significance of 8$\sigma$ in
integrated \hi flux is expected to result in a high degree of
completeness in our sample (cf. Rosenberg \& Schneider \cite{rose02}),
this can also be tested by evaluating the average value of V/V$_{max}$
(Schmidt \cite{schm68}). For a complete sample of a uniform density
volume we expect $<$V/V$_{max}>$~=~0.5.  In the absence of density
corrections we actually find $<$V/V$_{max}>$~=~0.35 for our sample, while
after discrete density correction we find $<$V/V$_{eff}^\prime>$~=~0.57,
where D$_{max}$ is based on the local 8$\sigma$ limit (which varies
with distance as in Fig.~\ref{fig:specrms}) over a linewidth of
1.5$\times$W$_{20}$. If instead we define D$_{max}$ by the local
8$\sigma$ limit over a linewidth of 1.2$\times$W$_{20}$ we obtain
$<$V/V$_{max}>$~=~0.31 and $<$V/V$_{eff}^\prime>$~=~0.50. This may suggest
that we have been somewhat too conservative in assessing significance
based on the larger velocity intervals. However, a roll-off in
completeness at the lowest significance levels can also result in an
elevated expectation value of $<$V/V$_{eff}>$ (Rosenberg \& Schneider
\cite{rose02}). Our value of $<$V/V$_{eff}^\prime>$~=~0.57 is comparable
to that found by Rosenberg \& Schneider (\cite{rose02}) for the ADBS
$<$V/V$_{eff}>$~=~0.60. 

\section{Summary and Discussion}
\label{sec:discussion}

Our unbiased \hi survey of 1800 deg$^2$ in the northern sky has allowed
recognition of a number of significant points regarding the \hi
content, distribution and environment of nearby galaxies. From the
analysis of some 500 candidate detections we have extracted a moderately
complete sample of 155 galaxies (listed in Table~\ref{tab:higal} and
illustrated in Fig.~\ref{fig:lgcs}) with an
integrated \hi flux in excess of 8$\sigma$ at distances between 5 and
80~Mpc. Seven of the detections occur so near the boundary of our two
segments of velocity coverage (near V$_{Hel}$~=~2800 \kms), and two so
near the edges of our spatial coverage, that their derived parameters
are unreliable (although the detections themselves are secure). This
leaves 146 detections with derived parameters of high quality. A
plausible optical galaxy ID was found within a 30 arcmin search radius
for all but one of the 8$\sigma$ detections, although one object was
previously uncataloged and three others had no previous red-shift
determination. Twenty-three objects (or their uncataloged companions)
are detected in \hi for the first time.

We have characterized the environment of each detected galaxy by
performing a search within NED for all cataloged objects within a
radius of 30 arcmin (corresponding to a possible contribution within
our telescope beam). These (potential) companions have been tabulated
in three categories in Table~\ref{tab:higal}, namely; (a) confused for
objects within 400~\kms\ of the primary ID, (b) unconfused for objects
offset by 400 to 1000~\kms\ from the primary ID, and (c) possibly
confused for objects of unknown red-shift. It will remain difficult to
assess the actual liklihood of association for objects in this last
category, until red-shift determinations become available. For the
moment we will regard only those objects with entries in catgory (a) as
``confused'' and all others as ``unconfused''.

We determine agreement of our absolute flux scale to the weighted
average of all previous determinations of the \hi flux (as tabulated by
LEDA) to better than 1\% for our unconfused detections, as shown in
Fig.~\ref{fig:flxflx}. Confused objects show a systematic excess \hi
flux in our large survey beam.

\subsection{Centroid Offsets of Gas and Stars}

Since our survey was not targeted at known galaxies, we have an
independent determination of the position centroid for each detected
object. The majority of apparent offsets between the gaseous and
stellar distributions are consistent with the substantial uncertainties
that follow from a large survey beam and only moderate signal-to-noise,
as shown in Fig.~\ref{fig:offset}. However, a number of significant
centroid offsets (greater than 5$\sigma$) are detected in nominally
unconfused galaxies which are indicated in Table~\ref{tab:higal} by
entering the symbol ``o'', ``+'' or ``++'' in the Note column. These
have been divided somewhat arbitrarily into two categories, depending
on whether the linear centroid offset is less than 10~kpc (category
``o'') or greater than 10~kpc (categories ``+'' and ``++''). The
reasoning behind this division is that a 10~kpc limit may mark a
plausible distinction between internal asymmetries of individual
objects and the larger scales that are more likely to indicate external
gaseous components.

Determining the cause of these significant centroid offsets requires
higher resolution imaging. One extensive source of high resolution
imaging is the WHISP survey (Kamphuis et al. \cite{kamp96},
http://www.astro.rug.nl/\~ whisp) which has targeted some 200 UGC
galaxies north of Dec~=~20$^\circ$ with synthesis observations using
the WSRT array. WHISP observations are currently available for only a
small fraction of the 155 galaxies in our 8$\sigma$ sample. The WHISP
results for each galaxy are summarized in a web-accessible data
overview consisting of a series of images of the integrated \hi
distribution and accompanying velocity field at three different angular
resolutions, of about 15, 30 and 60~arcsec, together with a global \hi
profile, a major axis position-velocity plot and an optical reference
image. The most relevant component of this overview for our purposes is
the distribution of integrated \hi at the lowest angular resolution of
60~arcsec, where the highest surface brightness sensitivity is reached.
For the four of 11 instances of significant centroid offset noted in
Table~\ref{tab:higal} that have already been imaged in the WHISP survey
we comment briefly on what is seen in the 60~arcsec integrated \hi
image:

\begin{itemize}
\item NGC~7640~o : asymmetric with extensions. 
\item UGC~12732~o : possible companions.
\item UGC~731~o : asymmetric.
\item NGC~925~o : asymmetric with extensions.
\end{itemize}

All four cases that have been imaged with high resolution show
large-scale asymmetries or possible uncataloged companions. Our
tentative conclusion is that our measured centroid offsets are indeed
indications of substantial asymmetries and the presence of possible
uncataloged gas-rich companions in the immediate vicinity of the
primary ID.

\subsection{Uncataloged companions}

The issue of uncataloged gas-rich companions is also addressed by the
comparison of our survey flux densities with those measured previously
for nominally unconfused galaxies in Figs.~\ref{fig:ratdm} and
\ref{fig:ratdma}. Although the comparison with the heterogenous LEDA
data has a large degree of scatter, the comparison with the Arecibo
data may indicate an excess of \hi at large radii.

The Arecibo data were taken from the
Pisces-Perseus supercluster survey (Giovanelli \& Haynes \cite{giov85},
Giovanelli et al. \cite{giov86}, Giovanelli \& Haynes \cite{giov89},
Wegner et al. \cite{wegn93} and Giovanelli \& Haynes \cite{giov93}). We
plot ratios of both the {\it observed\ } and the {\it corrected\ } \hi
flux density. In the latter case approximate corrections for telescope
pointing errors and a model of the galaxy extent were applied to the
Arecibo data. These data were only available for 20 of our unconfused
detections, which vary in distance from about 7 to 70~Mpc. The Arecibo
beam (3.3 arcmin FWHM at the time of those observations) has a linear
dimension that varies from about 7 to 70~kpc over the distance range
above, while our survey beam varies from about 100~kpc to 1~Mpc in
diameter.  Four of the unconfused galaxies have {\it observed\ } flux
ratios, $R$, of unity within our 1$\sigma$ errors, including the
nearest object. All of the other galaxies have an excess detected \hi
flux in our larger survey beam which varies from about 10\% to
300\%. The comparison of {\it corrected\ } flux ratios is more
ambiguous, since at least five of the data-points now have apparent
flux ratios which are significantly less than unity. As indicated in
\S~\ref{sec:results} above, each of our survey detections was examined
for evidence of source resolution effects and our absolute flux scale
is well-defined, making such apparent deficits in our detected \hi flux
difficult to understand. Given the approximate nature of the
corrections applied to the observed Arecibo fluxes it seems likely that
they may (on occasion) result in a degree of over-compensation for
missed flux.

The straightforward conclusion that can be drawn from the {\it
observed\ } flux ratio plots, it that in most cases, the \hi
distribution must be significantly more spatially extended than the
Arecibo beam, even when this beam subtends 50--70~kpc.

High resolution imaging will be necessary to determine, on a
case-by-case basis, where the excess detected flux actually
resides. WHISP data (http://www.astro.rug.nl/\~ whisp) are currently
available for 7 of the 20 unconfused galaxies that have Arecibo flux
measurements. Inspection of the 60~arcsec integrated \hi image in the
WHISP database yields the following assessment:

\begin{itemize}
\item NGC~7286 :  $R$=1.79, possible extensions.
\item UGC~12693 : $R$=1.33, possible extensions.
\item UGC~12713 : $R$=1.05, asymmetric.
\item UGC~1856 :  $R$=1.15, nothing unusual.
\item NGC~972 : $R$=1.38, asymmetric, extensions, possible companions.
\item NGC~1012 :  $R$=1.21, 10\% of \hi flux in uncataloged companion.
\item NGC~1056 :  $R$=1.72, extensions.
\end{itemize}

The two cases showing the largest excess \hi flux in the WSRT survey
beam are notable for having possible extensions at low surface
brightness in the distribution of integrated \hi in the WHISP data. 
Only in one of these seven cases, has a distinct uncataloged companion been
detected in the WHISP imaging. Deeper imaging of a larger field-of-view
will be needed to search for additional uncataloged companions to
account for the excess detected \hi flux.

An intriguing possibility for the location of the excess detected \hi
flux is that it resides in a relatively diffuse distribution subtending
a few 100~kpc in the vicinity of the primary target. This is exactly
the type of hypothetical distribution, in the environment of M31, which
motivated the negative velocity component of our wide-field \hi survey.
Such distributions were found (De Heij et al. \cite{dehe02}) to provide
the best-fit to the spatial and kinematic distribution of the compact
high--velocity cloud population in the vicinity of the Galaxy. The
best-fitting models of this type consist of gas bound to low-mass
dark-matter halos with a steep power-law ($\alpha=-1.7$) distribution
in number as function of neutral gas mass and are concentrated around their
major galaxy host in a Gaussian distribution with a spatial dispersion
of 150--200~kpc. The total \hi mass predicted in these Local Group
models to survive (ram-pressure- and tidal-stripping) to the present
day amounts to some 1.2$\times10^9$M$_\odot$. Compared to the
8$\times10^9$M$_\odot$ of \hi in M31 and the Galaxy, this corresponds
to an excess \hi mass of about 15\% distributed on scales of a few
hundred kpc. Only a handful of the rare, massive components might be
identifiable as discrete objects, while the rest of the distribution
would merely contribute to a diffuse enhancement of \hi mass, centered
on the systemic velocity of the host.

This possibility can and should be tested with a dedicated experiment.

\subsection{Spatial Variance of the HIMF }

In \S \ref{sec:results} we derive an HIMF from our 8$\sigma$ sample of
background galaxies. Despite the fact that our survey region covered
some 1800 deg$^2$ and therefore sampled a variety of environments along
the line-of-sight, it is impossible to overcome the fact that in a
flux-density limited sample, all low mass detections must of necessity
be very nearby. In consequence, the power-law slope and normalization
of the HIMF are strongly dependent on whether or not there are
significant variations of the average galaxy density with distance. We
have quantified such a variation by utilizing cataloged optical
galaxies in our survey volume as shown in Figs.~\ref{fig:lfdist} and
\ref{fig:lgsover}. The absolute density of optical galaxies as function
of distance was determined by fitting for the normalization of the
``standard'' luminosity function determined by Norberg et
al. (\cite{norb02}) in a sequence of heavily over-lapping
sub-samples.  Density variations with distance of more than an order of
magnitude are derived within our survey volume. Accounting for this
variation leads to very substantial changes in the best-fitting HIMF
parameters, as seen in Fig.~\ref{fig:lgshimf}. Before correction we
obtain log(M$_*$)~=~10.15$\pm0.1$, $\Theta_*$~=~9.5$\pm3\times10^{-4}$
and $\alpha$~=~$-$1.5$\pm0.1$, and after correction
log(M$_*$)~=~9.85$\pm0.07$, $\Theta_*$~=~55$\pm15\times10^{-4}$ and
$\alpha$~=~$-$1.28$\pm0.1$, where the error estimates come from the
2$\sigma$ contours of $\Delta\chi^2$ in the fit parameters shown in
Fig.~\ref{fig:chipar}.

Only after applying the density correction is statistical agreement
realized with the HIPASS BGC values (Zwaan et al. \cite{zwaa03})
log(M$_*$)~=~9.79$\pm0.06$, $\Theta_*$~=~86$\pm 21\times10^{-4}$ and
$\alpha$~=~$-$1.30$\pm0.08$ and to a lesser extent with the Arecibo
Dual-Beam Survey (ADBS) values (Rosenberg \& Schneider \cite{rose02})
log(M$_*$)~=~9.88, $\Theta_*$~=~58$\times10^{-4}$ and
$\alpha$~=~$-$1.53.

Schneider et al. (\cite{schn98}) and Rosenberg \& Schneider
(\cite{rose02}) also consider the variation of number density of
optical galaxies in the ADBS survey region.  However they determine
only a relative, rather than an absolute density and do so on the basis
of galaxy number counts rather than explicit fitting to the complete
portion of the luminosity function.  They conclude that density
corrections have only a minor impact on the form and normalization of
the HIMF in their sample, although their normalization does increase
from 48 to 58$\times10^{-4}$ (Mpc$^{-3}$dex$^{-1}$), when relative
density corrections are applied. These authors also make use of the
``integral'' formulation of density correction discussed in \S
\ref{sec:results} rather than the ``discrete'' formulation which we
find leads to substantially reduced fit residuals.

The absolute normalization of the HIPASS BGC HIMF is not a trivial
procedure (Zwaan et al. \cite{zwaa03}). Since a maximum likelihood
method has been employed, the shape if the HIMF should be
well-determined, but only in terms of a relative density. The
normalization is determined after the fact by carrying out integrations
of the derived selection function. Since a reasonable range of detected
masses in the HIPASS sample is only achieved inside of about 25~Mpc,
the normalization is, of necessity, also tied to this distance range.
To assess the impact of a possible variation of galaxy density with
distance within the HIPASS sample, we have also fit for the optical
galaxy normalization using all LEDA galaxies at $\delta<0$, just as in
\S \ref{sec:results}, for our own survey volume.  Given the larger
optical galaxy sample size, we defined 25 overlapping sub-samples with
sample populations varying linearly from a minumum of 100 galaxies at
the nearest distances to 1500 galaxies at the maximum distance.  This
distribution is plotted as the dotted line in
Fig.\ref{fig:lgsover}. The southern hemisphere has a galaxy density
which is equal to the Norberg et al. (\cite{norb02}) value between
about 8 and 25 ~Mpc. At larger distances there appears to be a smooth
decline to about 30\% of this density by 80~Mpc. At distances smaller
than about 8~Mpc there also appears to be a decline in galaxy density
to about 50\% of nominal. It seems that over the critical distance
range of 10--25~Mpc, the galaxy density within the HIPASS survey volume
is essentially the nominal one, suggesting that the HIPASS BGC value of
$\Theta_*$~=~86$\pm 21\times10^{-4}$ Mpc$^{-3}$dex$^{-1}$ should be
quite reliable. The apparent down-turn in galaxy density below about
8~Mpc may have some consequence for the apparent shape of the HIPASS
BGC HIMF below log(M)~=~7.5, since such systems could only be detected
out to 8~Mpc in the HIPASS data.

For comparison we have also determined and plotted the optical galaxy
density in the same way for the entire northern hemisphere in
Fig.\ref{fig:lgsover}. The northern hemisphere distribution is quite
different than the southern, with a moderate over-density (about 50 \%)
inside of 15~Mpc, followed by a relative dearth of galaxies between 20
and 60~Mpc. It seems quite conceivable that the apparent discrepancies
between the HIPASS BGC and the ADBS HIMF parameters may be a
consequence of such large-scale differences in the galaxy distribution.

\begin{acknowledgements}
We acknowledge useful discussions and feed-back on methods with
L. Staveley-Smith and M. Zwaan.
The Westerbork Synthesis Radio Telescope is operated by the Netherlands
Foundation for Research in Astronomy under contract with the
Netherlands Organization for Scientific Research. We have made use of
the LEDA database (http://leda.univ-lyon1.fr). This research has made
use of the NASA/IPAC Extragalactic Database (NED) which is operated by
the Jet Propulsion Laboratory, California Institute of Technology,
under contract with the National Aeronautics and Space Administration.
The Digitized Sky Surveys were produced at the Space Telescope Science
Institute under U.S. Government grant NAG W-2166. The images of these
surveys are based on photographic data obtained using the Oschin
Schmidt Telescope on Palomar Mountain and the UK Schmidt Telescope. The
plates were processed into the present compressed digital form with the
permission of these institutions.

\end{acknowledgements}

\end{document}